\documentclass[american,preprint]{aastex631}
\usepackage[T1]{fontenc}
\usepackage[utf8]{inputenc}
\usepackage{color}
\usepackage{amstext}
\usepackage{amssymb}
\usepackage{graphicx}
\usepackage{wasysym}

\makeatletter

\providecommand{\tabularnewline}{\\}

\submitjournal{ApJ}
\received{May 15, 2026}
\revised{June 21, 2026}
\accepted{July 8, 2026}

\usepackage{amsmath}
\newcommand\cemda{CeMDA}
\shorttitle{Circumbinary approach for spin-orbit resonances}
\shortauthors{F. Roig}

\makeatother

\usepackage{babel}
\begin{document}

\title{A circumbinary approach to the study of spin-orbit resonances around
irregular shaped bodies. Application to the Quaoar
system.}

\author[0000-0001-7059-5116]{Fernando Roig}
\affiliation{Observatório Nacional, Rio de Janeiro, RJ 20921-400, Brazil}
\begin{abstract}
We propose to model spin-orbit resonances that appear in ring systems
around minor bodies of the Solar System using a circumbinary approximation.
In our model, the ellipsoidal/irregular shape of the minor body is
replaced by a binary dumbbell, i.e., two equal masses evolving in
circular orbits around their center of mass. This allows us to apply
the equations of motion of the restricted circumbinary $N$-body problem,
duly adjusted to mimic the rotation of the central body and its quadruple
momentum. The equations also allow for the simple inclusion of other
perturbing bodies, like small satellites, enabling the analysis of
the simultaneous effect of spin-orbit resonances and mean motion resonances
on the ring dynamics. The goal of the circumbinary model is to substitute
the study of a given spin-orbit resonance by a surrogate mean motion
resonance, allowing for the application of well established numerical
and semianalytical models to map the topology and stability of these
resonances. We discuss the differences between the circumbinary model
and a triaxial ellipsoid. The model is also extended to study the
problem of a mass anomaly. We present some applications to the dynamics
of Quaoar's ring system, indicating that spin-orbit resonances do
not seem to play any relevant role in the present dynamics of the
ring particles.
\end{abstract}

\keywords{Trans-Neptunian objects (1705) -- Spin-orbit resonances (2296) --
Ring resonance (2295) -- Celestial mechanics (211)}

\section{Introduction}\label{sec:Introduction}

Spin-orbit resonances (SORs) occur when the spin frequency $\nu_{i}$
of a body in a given dynamical system is commensurable with the orbital
mean motion $n_{j}$ of another body. In the context of a two-body
system, the orbital mean motion of both bodies is the same, and that
leads to the usual view that a SOR involves a commensurability between
the rotation and the translation periods of the same body.

SORs are ubiquitous in the Solar System \citep{1966AJ.....71..425G,1967AJ.....72..662G}
as well as in exoplanetary systems \citep{2019AA...630A.102C}, and
most of them are driven by tidal forces. The tidal deformation caused
by a mass on an extended body dissipates energy and drives the system
to an equilibrium, where the bulge of the extended body tends to align
with the direction of the perturber. This induces a synchronization
between the rotation and the translation. The classical example is
that of the Earth-Moon system, where the Moon is in a 1:1 SOR. Although
this SOR is the most common to attain in the framework of tidal perturbations,
other are possible like the 3:2 SOR in the Sun-Mercury system, which
is driven not only by the solar tide but also by the high eccentricity
of Mercury's orbit.

Assuming that $\nu_{i}\geq n_{j}$, the resonant angle of a $k$:$l$
SOR may be generically defined as
\begin{equation}
\sigma=k\lambda_{j}-l\psi_{i}-(k-l)\varpi_{j}\label{eq:reso_angle}
\end{equation}
where $\psi_{i}=\nu_{i}t+\psi_{0,i}$ is the rotational phase of the
extended body, $\lambda_{j},\varpi_{j}$ are the mean longitude and
the longitude of pericenter of the orbit of the other body, $k,l$
are integers, and all the angles are measured with respect to the
same fixed direction. In this context, $l$ is referred to as the
degree of the resonance, while $q=k-l$ is referred to as the order
of the resonance. The rotational phase, in this case, is defined as
the angle described by the axis of less moment of inertia of the extended
body. In the case of tidal SORs, the theory predicts that the resonant
angle librates around an equilibrium point $\sigma_{0}$. For the
Earth-Moon system, $\sigma_{0}=0,\pi$, which implies $\psi_{\leftmoon}=\lambda_{\oplus}$
or $\psi_{\leftmoon}+\pi=\lambda_{\oplus}$, so the lunar bulge is
always aligned towards the Earth direction. For the Sun-Mercury system,
$\sigma_{0}=0$ which implies $2\psi_{\mercury}=3\lambda_{\odot}-\varpi_{\odot}$,
so Mercury's bulge always points towards the Sun at perihelion \citep{2004AA...413..381R}.

However, SORs may also appear naturally in a system, without being
necessarily driven by tidal mechanisms. This is the case when a test
particle or a small mass body evolves around a rotating body of irregular
shape, and only the gravitational forces arise. Examples are the geostationary
satellites around the Earth, and the ring systems around Centaurs
and Trans-Neptunian Objects (TNOs). When the central body is deformed
along its rotational axis, like in the polar oblateness, the effect
of this kind of SORs is, in principle, negligible due to the azimuthat
symmetry of the potential. But when the body is deformed along the
equator, like in a prolate shape, the effects of the SOR may be relevant
due to the lack of rotational symmetry. 

This kind o SOR model, where the spin-orbit phase
lock is caused by an external rotating potential forcing, is similar
to the so-called conservative spin-orbit model \citep{2000CeMDA..76..229C,2007PSS...55..889C},
where a free rotating ellipsoid of mass $m$ orbits around a massive
body without accounting for the tidal lag. However, there is a subtle
difference between the two models. In the latter case, that we could
refer to as the ``classical'' SOR (because this is the SOR model
that is usually addressed in almost all the literature on spin-orbit
dynamics), the dynamical variable is the rotational phase $\psi$,
while the orbital phase $\lambda$ acts as the time dependence of
the potential. In other words, the dynamics is ruled by a Hamiltonian
of the form
\begin{equation}
\mathcal{H}(\psi,p_{\psi},\lambda,\Lambda)=\frac{p_{\psi}^{2}}{2}+n\Lambda+\mathcal{R}(\psi,\lambda)
\end{equation}
where 
\begin{equation}
\mathcal{R}(\psi,\lambda)\propto-\left(\frac{a}{r}\right)^{3}\cos\left(2\psi-2f\right),\label{eq:rcsor}
\end{equation}
$p_{\psi}=\dot{\psi}$, $a$ is the orbital semi-major axis, $r$
is the orbital radius, and $f$ is the true anomaly. On the other
hand, for the kind of SOR that we are interested here, the dynamical
variable is the orbital phase, while the rotational phase acts as
the time dependence of the potential. In other words, the dynamics
is ruled by a Hamiltonian of the form
\begin{equation}
\mathcal{H}(\lambda,L,\psi,\Psi)=-\frac{G^{2}m^{2}}{2L^{2}}+\nu\Psi+\mathcal{R}(\lambda,\psi)
\end{equation}
where 
\begin{equation}
\mathcal{R}(\lambda,\psi)\propto-\left(\frac{a}{r}\right)^{3}\cos\left(2f-2\psi\right),\label{eq:rfsor}
\end{equation}
and $L=\sqrt{Gma}$. Note that in both kind of SORs, the disturbing
function has the same form, and thus the same harmonic structure in
terms of $\lambda,\psi$. However, the main unperturbed parts have
opposite sign. This subtlety will have consequences on the stability
of the equilibria of the systems, as we will see later.

Here, we are interested in analyzing the SORs with a specific application
to the orbital environment around 5000 Quaoar. This
is a TNO that was recently identified as having, at least, two rings
located at 2,500 and 4,000 km from the central body, respectively.
Due to the estimated mass and dimensions of Quaoar, both
rings appear to be located beyond the Roche limit ($\sim1,800$ km),
raising questions about their survival without accreting to form satellites
\citep{2023Natur.614..239M,2023AA...673L...4P}. Quaoar also has a
satellite called Weywot, located about 13,300 km away \citep{2025RSPTA.38340200B},
and it is also discussed the existence of a second, much less massive,
moon orbiting at about 5,800 km \citep{2025ApJ...993L..38P}. In this
context, the outermost ring appears to be confined by two resonances:
on its inner edge, by the 6:1 mean motion resonance (MMR) with Weywot
\citep{2023MNRAS.525.3376R} and, on its outer edge, by the 3:1 SOR
with the rotational period of Quaoar \citep{2025AA...704A..23S}.
The latest estimates attribute to Quaoar the shape of a triaxial ellipsoid,
with axes $R_{x}=583.3$ km, $R_{y}=555.3$ km, and $R_{z}=510.0$
km, rotating about its minor axis \citep{Margoti2024}. 

In this work, we propose to model the SORs using a circumbinary approximation,
in which the ellipsoidal shape of Quaoar is replaced by a binary dumbbell,
i.e., two equal masses $m_{1},m_{2}$ following circular orbits around
their center of mass. Our applications will be focused
on the region beyond the estimated Roche limit. The main advantage
of this approach is to replace the perturbation generated by an extended
body, that require to expand the potential in spherical harmonics,
by the perturbation generated by point masses. This allows us to reduce
the study of a SOR to the study of a two-body MMR, and to apply well
established theoretical and numerical tools of MMRs. The approach
also allows us to simulate the evolution of the system using a pure
$N$-body algorithm, and to straightforwardly include the perturbations
of other bodies in the system, like satellites, enabling the analysis
of the simultaneous effect of the SOR and other MMRs around
Quaoar. Additionally, the circumbinary model can be easily generalized
to study the case of a mass anomaly configuration, by setting $m_{1}>m_{2}$,
that would be extremely cumbersome to treat using spherical harmonics
expansions (even though such a configuration goes
beyond the specific case of Quaoar).

On the other hand, the main disadvantages are that the model cannot
mimic the triaxiality of the central body, being restricted to the
case of a purely prolate shape, nor it can simulate a body rotating
around an arbitrary axis not coincident with the axis of maximum moment
of inertia.

The idea of simulating a prolate ellipsoid by two pseudo-Keplerian
masses has been recently explored by \citet{2025AA...702L..15B} to
study the confinement of narrow rings around minor bodies in a non
conservative framework, but the authors did not focus on the analysis
of the SORs. \citet{2022MNRAS.510.1450M} explored
the case of a spherical body with a mass anomaly, but, although they
did focus on the description of the SORs, they used a significantly
different technique (Poincare surface of section), and they did not
apply their model to the case of Quaoar. The mass anomaly model has
also been investigated more recently by \citet{2026A&A...708A.304B},
through an semi-analytical modeling aiming to describe the topology
of the SORs. Their work also focus on the problem of resonance capture,
and considered a Quaoar-mass central body with a mass anomaly corresponding
to a mass ration of 0.01. Finally, the analysis of SORs around a prolate
rotating ellipsoid has been addressed by \citet{2023MNRAS.525...44R},
also using the Poincare surface of section technique, but again without
applications to Quaoar.

At the time of submitting this work, we became aware of a similar
study by Gianuzzi et al. (2026, in preparation), which also analyzes
the dynamics of a particle around an irregular body represented by
a spherical object with two mass anomalies (boulders). Although the
model and methodology used by those authors differ from that applied
in this work, both studies present some results in common and can
be considered, in a way, complementary. We will refer to the work
of Gianuzzi et al. throughout this article when pertinent.

This paper is organized as follows: in Sect. \ref{sec:Model} we present
the circumbinary model, considering a pure numerical $N$-body algorithm
and a semianalytical model, and we discuss the main differences with
other models, like a triaxial ellipsoid or a mass anomaly. In Sect.
\ref{sec:Results} we present the application of the model to the
Quaoar system, first analyzing the case of an equal mass binary and
then the case of a mass anomaly. Finally, Sect. \ref{sec:Conclusions}
is devoted to the conclusions, and some specific
topics are treated in the Appendix.

\section{Model}\label{sec:Model}

\subsection{Numerical approach}\label{subsec:Numerical-approach}

Consider a hierarchical system of point masses $m_{1}\geq m_{2}\gg m_{3},\ldots,m_{N}$,
with barycentric positions $\vec{\rho}_{i}$ and momenta $\vec{\pi_{i}}=m_{i}\vec{\upsilon}_{i}$.
Define new positions such that:

\begin{equation}
\vec{r}_{1}=0;\qquad\vec{r}_{2}=\vec{\rho}_{2}-\vec{\rho}_{1};\qquad\vec{r}_{i}=\vec{\rho}_{i}-\frac{m_{1}\vec{\rho}_{1}+m_{2}\vec{\rho}_{2}}{m_{b}},\qquad i=3,\ldots,N
\end{equation}
with $m_{b}=m_{1}+m_{2}$ and $\vec{r}_{i}\gg\vec{r}_{1},\vec{r}_{2}$.
Therefore, the position of $m_{2}$ becomes referred to $m_{1}$ (bodycentric)
while the positions of $m_{i}$ are referred to the barycenter of
the binary pair $m_{1},m_{2}$ (binarycentric). The conjugated momenta
are:
\begin{equation}
\vec{p}_{1}=0;\qquad\vec{p}_{2}=\frac{m_{1}\vec{\pi}_{2}-m_{2}\vec{\pi}_{1}}{m_{b}}=\mu_{2}\vec{v}_{2};\qquad\vec{p}_{i}=\vec{\pi_{i}}=\mu_{i}\vec{v}_{i}
\end{equation}
with $\mu_{2}={\displaystyle \frac{m_{1}m_{2}}{m_{b}}}$ and $\mu_{i}={\displaystyle \frac{m_{b}m_{i}}{m_{b}+m_{i}}}$,
and the velocities are:
\begin{equation}
\vec{v}_{2}=\vec{\upsilon}_{2}-\vec{\upsilon}_{1};\qquad\vec{v}_{i}=\frac{m_{b}+m_{i}}{m_{b}}\vec{\upsilon}_{i}.
\end{equation}
Therefore, the velocity of $m_{2}$ is bodycentric, while the velocities
of $m_{i}$ are barycentric. In this context, the Hamiltonian of the
full $N$-body system is:
\begin{align}
\mathcal{H} & =\left(\frac{p_{2}^{2}}{2\mu_{2}}-\frac{Gm_{b}\mu_{2}}{r_{2}}\right)+\sum_{i=3}^{N}\left(\frac{p_{i}^{2}}{2\mu_{i}}-\frac{G(m_{b}+m_{i})\mu_{i}}{r_{i}}\right)\nonumber \\
 & +\sum_{i=3}^{N-1}\sum_{j=i+1}^{N}\left[-\frac{Gm_{i}m_{j}}{\left|\vec{r}_{i}-\vec{r}_{j}\right|}+\frac{1}{m_{b}}\vec{p}_{i}\cdot\vec{p}_{j}\right]\nonumber \\
 & +\,\sum_{i=3}^{N}Gm_{i}\left(\frac{m_{b}}{r_{i}}-\frac{m_{1}}{\left|\vec{r}_{i}+{\displaystyle \frac{m_{2}}{m_{b}}}\vec{r}_{2}\right|}-\frac{m_{2}}{\left|\vec{r}_{i}-{\displaystyle \frac{m_{1}}{m_{b}}}\vec{r}_{2}\right|}\right)\label{eq:hamil_full}
\end{align}
The first two terms represents the Keplerian motion of the binary
and the massive bodies, respectively, the third term is the mutual
perturbation among the massive bodies, and the last term is the mutual
perturbation between the massive bodies and the binary. This $N$-body
Hamiltonian is suitable to be solved using a second-order symplectic
integrator, like the one proposed by \citet{2010ASSL..366..239C}.

Now, we will consider the Hamiltonian above in the framework of a
restricted $N$-body problem, where:
\begin{itemize}
\item $m_{N}=0$ is a test particle, and
\item $m_{1},m_{2}$ do not feel the gravitational pull of the other bodies
in the system.
\end{itemize}
Then, the motion of the system becomes ruled by the following set
of Hamiltonians:
\begin{align}
\mathcal{H}_{\mathrm{bin}}(\vec{r}_{2},\vec{p}_{2}) & =\frac{p_{2}^{2}}{2\mu_{2}}-\frac{Gm_{b}\mu_{2}}{r_{2}}\label{eq:hamil_bin}\\
\mathcal{H}_{\mathrm{sat}}(\vec{r}_{i},\vec{p}_{i},t) & =\sum_{i=3}^{N-1}\left(\frac{p_{i}^{2}}{2\mu_{i}}-\frac{G(m_{b}+m_{i})\mu_{i}}{r_{i}}\right)+\sum_{i=3}^{N-2}\sum_{j=i+1}^{N-1}\left[-\frac{Gm_{i}m_{j}}{\left|\vec{r}_{i}-\vec{r}_{j}\right|}+\frac{1}{m_{b}}\vec{p}_{i}\cdot\vec{p}_{j}\right]\nonumber \\
 & +\sum_{i=3}^{N-1}Gm_{i}\left[\frac{m_{b}}{r_{i}}-\frac{m_{1}}{\left|\vec{r}_{i}+{\displaystyle \frac{m_{2}}{m_{b}}}\vec{r}_{2}(t)\right|}-\frac{m_{2}}{\left|\vec{r}_{i}-{\displaystyle \frac{m_{1}}{m_{b}}}\vec{r}_{2}(t)\right|}\right]\qquad\qquad i=3,\ldots,N-1\label{eq:hamil_sat}\\
\mathcal{H}_{\mathrm{ring}}(\vec{r}_{N},\vec{p}_{N},t) & =\frac{p_{N}^{\,2}}{2}-\frac{Gm_{b}}{r_{N}}+\sum_{i=3}^{N-1}\left[-\frac{Gm_{i}}{\left|\vec{r}_{N}-\vec{r}_{i}(t)\right|}+\frac{1}{m_{b}}\vec{p}_{N}\cdot\vec{p}_{i}(t)\right]\nonumber \\
 & +\frac{Gm_{b}}{r_{N}}-\frac{Gm_{1}}{\left|\vec{r}_{N}+{\displaystyle \frac{m_{2}}{m_{b}}}\vec{r}_{2}(t)\right|}-\frac{Gm_{2}}{\left|\vec{r}_{N}-{\displaystyle \frac{m_{1}}{m_{b}}}\vec{r}_{2}(t)\right|}\label{eq:hamil_ring}
\end{align}
where $\mathcal{H}_{\mathrm{bin}}$ represents the Keplerian (unperturbed)
motion of the binary dumbbell, $\mathcal{H}_{\mathrm{sat}}$ represents
the motion of the other massive bodies in the system, and $\mathcal{H}_{\mathrm{ring}}$
is the Hamiltonian of the test particle. Note that both $\mathcal{H}_{\mathrm{sat}}$
and $\mathcal{H}_{\mathrm{ring}}$ are non autonomous systems. This
restricted $N$-body system can be solved with the same symplectic
algorithm used for the full $N$-body system, with little modifications.

In order for the binary dumbbell Hamiltonian, Eq. (\ref{eq:hamil_bin}),
to mimic a rotating prolate ellipsoid of radii $R_{x}>R_{y}=R_{z}$
and total mass $m_{b}$, the following conditions should be fulfilled:
\begin{enumerate}
\item $m_{2}$ equals $m_{1}$, and they evolve in circular orbits at distance
$d$ from their common center of mass,
\item the orbital mean motion of the binary, $n_{2}$, equals the rotation
frequency of the ellipsoid, $\nu_{2}$, and
\item the moment of inertia of the binary equals the moment of inertia of
the ellipsoid.
\end{enumerate}
Of course the last condition cannot be achieved because the ellipsoid
is an extended body while the binary dumbbell is not, so they have
different bulk distributions. However, we can approach to the problem
if we consider the symmetric trace-free (STF) quadrupole tensor of
both systems, that only carries information about the anisotropic
contribution of the shape. In the inertial frame, this has the form
\citep{Jackson1999}:
\begin{equation}
Q_{kl}=\mathcal{Q}\,(3u_{k}u_{l}-\delta_{kl})\label{eq:STF_quad}
\end{equation}
 where $\vec{u}=(\cos\nu_{2}t,\,\sin\nu_{2}t,\,0)\equiv(\cos n_{2}t,\,\sin n_{2}t,\,0)$
and
\begin{center}
\begin{tabular}{ll}
${\displaystyle \mathcal{Q}=\frac{m_{b}}{5}(R_{x}^{2}-R_{y}^{2})}\qquad\qquad$ & for the prolate ellipsoid\tabularnewline
${\displaystyle \mathcal{Q}=\frac{m_{2}a_{2}^{2}}{2}}$ & for the equal mass binary dumbbell\tabularnewline
\end{tabular}
\par\end{center}

\noindent with $a_{2}=2d$. This provides a direct way to setup the
bodycentric semimajor axis of the binary
\begin{equation}
a_{2}=\sqrt{\frac{4}{5}(R_{x}^{2}-R_{y}^{2})}\label{eq:calib_dist}
\end{equation}
and to guarantee that the potential generated by the binary is equivalent
to the potential generated by the ellipsoid, up to the quadrupole
term. The matching of the quadrupole tensor can be done in the body
fixed frame, instead of the inertial frame, with exactly the same
result provided that $R_{x}$ is aligned with $a_{2}$.

The second condition can be achieved in different ways. One possibility
would be to simply force a solution $\vec{r}_{2}(t)=(a_{2}\cos n_{2}t,\,a_{2}\sin n_{2}t,\,0)$
and to insert it into the equations of motion. Here, instead, we chose
to explicitly solve Eq. (\ref{eq:hamil_bin}) considering fake masses
$m_{1}^{\prime},m_{2}^{\prime}$ such that:
\begin{equation}
\mathcal{H}_{\mathrm{bin}}\rightarrow\mathcal{H}_{\mathrm{bin}}^{\prime}=\frac{p_{2}^{2}}{2\mu_{2}^{\prime}}-\frac{Gm_{b}^{\prime}\mu_{2}^{\prime}}{r_{2}}
\end{equation}
 with
\begin{equation}
m_{b}^{\prime}=\frac{n_{2}^{2}a_{2}^{3}}{G};\qquad m_{1,2}^{\prime}=m_{1,2}\frac{m_{b}^{\prime}}{m_{b}}
\end{equation}
and $n_{2}=\nu_{2}$. This latter approach, that we refer to as pseudo-Keplerian
motion, is naturally incorporated into the symplectic
$N$-body algorithm, again with little modifications
in the code\footnote{The solution is advanced using the Gauss $f$and
$g$ functions \citep{1962fcm..book.....D}, which for $e=0$ reduce
to a matrix rotation by and angle $n_{2}\Delta t$. In terms of computational
effort, there is no actual gain in using the fake masses approach.}. Note that the fake masses are used only to advance the solution
of Eq. (\ref{eq:hamil_bin}); to solve Eqs. (\ref{eq:hamil_sat})
and (\ref{eq:hamil_ring}) we still use the true masses.

A final consideration concerns the setup of the initial orbital phase
of the binary, i.e. the initial mean longitude $\lambda_{2}\equiv\psi_{2}$.
If we want the test particle to be close to a given SOR, we have to
set
\begin{equation}
l\lambda_{2}=k\lambda_{N}-(k-l)\varpi_{N}-\sigma_{0}\label{eq:lambda}
\end{equation}
where $\sigma_{0}$ is the equilibrium point of the resonant angle.
We discuss later how to determine $\sigma_{0}$ for any SOR.

\subsection{Semianalytical approach}

As previously stated, the advantage of reducing the study of a SOR
to the study of a surrogate MMR relies in the possibility of applying
analytical and semianalytical tools developed for MMRs. Here, we apply
the model of \citet{2020CeMDA.132....9G} and \citet{2021AA...646A.148G}
to study the topology of the SOR.

Our application of the model of Gallardo consists in numerically perform
a first order averaging of the Hamiltonian of the test particle evolving
under the perturbation of the binary only. Following Eq. (\ref{eq:hamil_ring}),
we consider $\mathcal{H}=\mathcal{K}+\mathcal{R}$ with:
\begin{align}
\mathcal{K}(\vec{r},\vec{p}) & =\frac{p^{\,2}}{2}-\frac{Gm_{b}}{r}\nonumber \\
\mathcal{R}(\vec{r},t) & =\frac{Gm_{b}}{r}-\frac{Gm_{1}}{\left|\vec{r}+{\displaystyle \frac{m_{2}}{m_{b}}}\vec{r}_{2}(t)\right|}-\frac{Gm_{2}}{\left|\vec{r}-{\displaystyle \frac{m_{1}}{m_{b}}}\vec{r}_{2}(t)\right|}\label{eq:distrubing}
\end{align}
where we dropped off the subindex $N$ for simplicity. If we extend
the phase space to get an autonomous system close to a $k$:$l$ $\mathrm{MMR\equiv FSOR}$,
we can write:
\begin{equation}
\mathcal{H}=\mathcal{F}(a)+\mathcal{R}(a,e,I,\varpi,\Omega,\,\lambda(\sigma,\lambda_{2}),\,\lambda_{2})
\end{equation}
where
\begin{equation}
\mathcal{F}(a)=-\frac{Gm_{b}}{2a}-\frac{l}{k}n_{2}\sqrt{Gm_{b}a}
\end{equation}
and $\lambda$ depends on $\lambda_{2}$ through the resonant angle
$\sigma$. Thus, the disturbing function is averaged as:
\begin{equation}
\overline{\mathcal{R}}(\sigma)=\frac{1}{2\pi k}\int_{0}^{2\pi k}\mathcal{R}(\lambda(\sigma,\lambda_{2}),\,\lambda_{2})\,d\lambda_{2}\label{eq:averager}
\end{equation}
where the average is performed over a full orbital period of the test
particle, assuming that, in that interval of time, only the mean longitudes
vary and the remaining orbital elements can be kept fixed. In particular,
$a$ is fixed at its resonant value
\begin{equation}
a_{0}=\left[\frac{k^{2}}{l^{2}}\frac{Gm_{b}}{n_{2}^{2}}\right]^{1/3}
\end{equation}

The resulting averaged Hamiltonian is a one degree-of-freedom system:
\begin{equation}
\overline{\mathcal{H}}(a,\sigma)=\mathcal{F}(a)+\overline{\mathcal{R}}(a_{0},\sigma)
\end{equation}
where the locations of the equilibrium points $\sigma_{0}$ are defined
by the condition
\begin{equation}
\frac{\partial\overline{\mathcal{R}}}{\partial\sigma}=0
\end{equation}
The equilibrium points are stable if
\begin{equation}
\mathcal{F}_{SS}\overline{\mathcal{R}}_{\sigma\sigma}>0\label{eq:equilib-1}
\end{equation}
where $S=\sqrt{Gm_{b}a}/k$ is the canonical momentum associated to
$\sigma$ and
\begin{equation}
\mathcal{F}_{SS}=\left.\frac{\partial^{2}\mathcal{F}}{\partial S^{2}}\right|_{a_{0}},\qquad\qquad\overline{\mathcal{R}}_{\sigma\sigma}=\left.\frac{\partial^{2}\overline{\mathcal{R}}}{\partial\sigma^{2}}\right|_{\sigma_{0}}\label{eq:equilib}
\end{equation}
therefore, since $\mathcal{F}_{SS}<0$, the stability condition reduces
to $\overline{\mathcal{R}}_{\sigma\sigma}<0$. 

The libration period at a stable equilibrium point is given by
\begin{equation}
\tau\simeq\frac{2\pi}{k}\frac{a_{0}}{\sqrt{3\left|\overline{\mathcal{R}}_{\sigma\sigma}\right|}}
\end{equation}
and the half-width of the resonance by
\begin{equation}
\Delta a\simeq\frac{1}{n}\sqrt{\frac{8}{3}\Delta\overline{\mathcal{R}}}\label{eq:widths}
\end{equation}
where $\Delta\overline{\mathcal{R}}=\overline{\mathcal{R}}_{\mathrm{max}}-\overline{\mathcal{R}}_{\mathrm{min}}$.
It is worth noting that the value of $\overline{\mathcal{R}}_{\mathrm{max}}$
always correspond to a stable equilibrium point, while the value of
$\overline{\mathcal{R}}_{\mathrm{min}}$ may correspond to an unstable
equilibrium point or to a collision of the system, when $\overline{\mathcal{R}}_{\mathrm{min}}\rightarrow-\infty$.
In this latter case, $\overline{\mathcal{R}}_{\mathrm{min}}$ shall
be computed by imposing a mutual distance threshold between the test
particle and the binary, which we set here to 0.1 Hill radii. The
topology of the resonance is recovered by plotting the level curves
of $\overline{\mathcal{H}}(a,\sigma)=\mathrm{const.}$

\subsection{Ellipsoid equivalence}

As described in Sect. \ref{subsec:Numerical-approach}, by applying
Eqs. (\ref{eq:STF_quad}) and (\ref{eq:calib_dist}), we can guarantee
that the binary dumbbell potential would be equivalent to the prolate
ellipsoid potential up to the quadrupole term. We may wonder, however,
how the potential and the corresponding accelerations differ at higher
order multipole terms \citep{Jackson1999}. 

Due to symmetry, it is straightforward to show that the octupole term
is zero in both cases. A non zero octupole only appears when the binary
dumbbell has unequal masses, or when the ellipsoid has a mass anomaly,
as we discuss below.

The next non zero multipole arises at the hexadecapole term. Let us
define $\ell=(R_{x}^{2}-R_{y}^{2})/R_{x}^{2}$, then the STF hexadecapole
tensor of the prolate ellipsoid scales as $\mathcal{G}\propto m_{b}R_{x}^{4}\ell$,
and the corresponding contribution to the potential is
\begin{equation}
\Phi_{4}\sim\frac{G\left|\mathcal{G}\right|}{r^{5}}\propto\frac{Gm_{b}}{r^{5}}R_{x}^{4}\ell
\end{equation}
On the other hand, the STF hexadecapole of the equal masses binary
dumbbell scales as $\mathcal{G}\propto m_{2}a_{2}^{4}$, and by matching
the quadrupole using Eq. (\ref{eq:calib_dist}), we have that $a_{2}\propto R_{x}\sqrt{\ell}$,
thus 
\begin{center}
\begin{tabular}{ll}
$\mathcal{G}\propto m_{b}R_{x}^{4}\ell$ & for the prolate ellipsoid\tabularnewline
$\mathcal{G}\propto m_{b}R_{x}^{4}\ell^{2}\qquad\qquad$ & for the matched equal mass binary dumbbell\tabularnewline
\end{tabular}
\par\end{center}

\noindent and the binary hexadecapole is suppressed by an extra factor
of $\ell$ relative to the ellipsoid. Therefore, the difference in
the potential at the hexadecapole term, for a test particle located
at a distance $r=\kappa R_{x}$, will be dominated by the ellipsoid
term:
\begin{equation}
\Delta\Phi_{4}\sim\frac{Gm_{b}}{R_{x}}\frac{\ell}{\kappa^{5}}
\end{equation}
and the relative error with respect to the monopole will be
\begin{equation}
\frac{\left|\Delta\Phi_{4}\right|}{\left|\Phi_{0}\right|}\sim\frac{\ell}{\kappa^{4}}\label{eq:error-hexa}
\end{equation}
Finally, the relative error in the accelerations (with respect to
the monopole) will also scale as $\ell/\kappa^{4}$. In the case of
Quaoar, using $R_{x}=583.3$ km and assuming a mean minor radius $R_{y}=532.6$
km, we have $\ell\simeq0.166$. For the inner ring, $\kappa\simeq4.5$,
which implies a relative acceleration error of $\sim4\times10^{-4}$,
while for the outer ring, $\kappa\simeq7.3$ and the relative acceleration
error is $\sim5.8\times10^{-5}$.

In the case of a triaxial ellipsoid rotating around its minor axis,
octupole is still zero, but it is no longer possible for the binary
dumbbell to match all the quadrupole STF tensor components at once.
Suppose that we choose to match the $Q_{xx}$ component of the tensor
and, for simplicity, let's work in the body fixed frame (the result
will be the same if working in the inertial frame). We have that,
for the binary, $Q_{xx}^{(b)}=-Q_{yy}^{(b)}/2=-Q_{zz}^{(b)}/2$, and
for the ellipsoid $Q_{xx}^{(e)}\neq Q_{yy}^{(e)}\neq Q_{zz}^{(e)}$,
therefore, since both tensors are traceless, we have: 
\begin{equation}
Q_{xx}^{(e)}=Q_{xx}^{(b)};\qquad Q_{yy}^{(e)}=Q_{yy}^{(b)}+\gamma;\qquad Q_{zz}^{(e)}=Q_{zz}^{(b)}-\gamma
\end{equation}
where
\begin{equation}
\gamma=\frac{Q_{yy}^{(e)}-Q_{zz}^{(e)}}{2}=\frac{3}{10}m_{b}(R_{y}^{2}-R_{z}^{2})
\end{equation}
is the mismatch parameter. Defining $\ell^{\prime}=(R_{y}^{2}-R_{z}^{2})/R_{y}^{2}$,
the unmatched part of the quadrupole potential at a distance $r=\kappa R_{y}$
will be
\begin{equation}
\Delta\Phi_{2}\sim\frac{G\left|\gamma\right|}{r^{3}}\propto\frac{Gm_{b}}{R_{y}}\frac{\ell^{\prime}}{\kappa^{3}},
\end{equation}
and the relative error with respect to the monopole will be
\begin{equation}
\frac{\left|\Delta\Phi_{2}\right|}{\left|\Phi_{0}\right|}\sim\frac{\ell^{\prime}}{\kappa^{2}}\label{eq:error-tri}
\end{equation}
with the same scaling for the accelerations. In the case of Quaoar,
using $R_{y}=555.3$ km and $R_{z}=510.0$ km, we have $\ell^{\prime}\simeq0.157$,
which for the inner ring implies a relative acceleration error of
$\sim7.7\times10^{-3}$, while for the outer ring an error $\sim3\times10^{-3}$.
These discrepancies are an order of magnitude larger than the ones
due to the hexadecapole mismatch.

Although these errors may be significant, it is not
clear whether this could actually affect the secular evolution of
the system. But we may expect that they do not interfere in the analysis
of the ``short term'' resonant dynamics. We will come back to this
issue in the Appendix.

\subsection{The mass anomaly case}

The study of bodies with a mass anomaly has caught the attention in
recent times as an alternative to study irregular
shaped bodies \citep{2022MNRAS.510.1450M}, and to explain the origin
and sustainability of narrow ring systems around minor bodies \citep{2025AA...702L..15B,2026A&A...708A.304B}.
Quaoar appears to have a well established ellipsoidal
shape, so considering a mass anomaly model does not constitute an
actual application to the Quaoar system. Nevertheless, here and later
in Sect. \ref{subsec:Mass-anomaly-system}, Quaoar will be used only
as an example to understand how and why the dynamics could change
when passing from an equal mass dumbbell to an unequal mass one.

Our circumbinary approach allows to easily simulate the mass anomaly
case, by setting $m_{1}>m_{2}$. In this case, matching the quadrupole
tensor of an ellipsoid (either prolate or triaxial) appears to be
meaningless, but it can still be done. In this case, Eq. (\ref{eq:STF_quad})
still holds, but now with
\begin{equation}
\mathcal{Q}=\frac{m_{1}m_{2}}{m_{b}}a_{2}^{2}
\end{equation}
and Eq. (\ref{eq:calib_dist}) becomes
\begin{equation}
a_{2}=\sqrt{\frac{1}{5}\frac{m_{b}^{2}}{m_{1}m_{2}}(R_{x}^{2}-R_{y}^{2})}\label{eq:dist_manomaly}
\end{equation}
which produces a dumbbell more separated than in the equal mass case.
By construction, this unequal mass dumbbell will generate a potential
and accelerations that are exactly the same of the equal mass case
(and the prolate ellipsoid), up to the quadrupole term. The main differences
arise at the octupole term, that is non zero in the unequal mass case.
This octupole introduces a contribution to the potential
of the order of
\begin{equation}
\Delta\Phi_{3}\equiv\Phi_{3}\sim\frac{G\left|S\right|}{r^{4}}
\end{equation}
where
\begin{equation}
S=\frac{m_{1}m_{2}}{m_{b}^{2}}(m_{2}-m_{1})a_{2}^{3}
\end{equation}
For a test particle at a distance $r=\kappa R_{x}$, the contribution
of the octupole relative to the monopole is then
\begin{equation}
\frac{\left|\Delta\Phi_{3}\right|}{\left|\Phi_{0}\right|}\sim\frac{m_{1}-m_{2}}{\sqrt{m_{1}m_{2}}}\frac{\ell^{3/2}}{\kappa^{3}}=\frac{1-\eta}{\sqrt{\eta}}\frac{\ell^{3/2}}{\kappa^{3}}\label{eq:erros_octupole}
\end{equation}
with $\eta=m_{2}/m_{1}$. The accelerations follow the same scaling
and, for a mass ratio $\eta=0.1$, gives values of $\sim2\times10^{-3}$
and $\sim5\times10^{-4}$ for the inner and outer rings of Quaoar,
respectively. Note that the smaller the mass ratio, the larger the
error, which seems to be counterintuitive. The reason is that, by
forcing to match the quadrupole, the separation of the binary may
increase indefinitely; indeed, for a mass ratio of $\eta\sim0.002$,
the mass anomaly would be located at the distance of the inner ring.

This issue also poses a particular limitation to the mass anomaly
model. If we set the small mass to be located at the surface of the
main body, this implies to set $a_{2}\simeq R_{x}$ and Eq. (\ref{eq:dist_manomaly})
becomes
\begin{equation}
\frac{m_{1}m_{2}}{m_{b}^{2}}\simeq\frac{\ell}{5}
\end{equation}
which for Quaoar gives $m_{2}/m_{b}\simeq0.034$. This means that,
assuming Quaoar as a spherical body, and putting a mass anomaly of
about 3\%-4\% of its mass on its surface, will produce a perturbation
that, at the quadrupole order, will be indistinguishable form the
perturbation of Quaoar's ellipsoid. And at the octupole order, according
to Eq. (\ref{eq:erros_octupole}), the relative contribution
in the accelerations will be only $\sim10^{-4}$ and $\sim3\times10^{-5}$
for the inner and outer ring, respectively.

If instead of matching the quadrupole tensor we keep the binary separation
fixed, we may wonder what is the difference introduced,
at the quadrupole level, by the mass anomaly with respect to the equal
mass dumbbell. Suppose that matching the equal mass dumbbell to the
ellipsoid quadrupole sets a separation $d_{2}$ for the binary (Eq.
\ref{eq:calib_dist}). Putting the unequal mass dumbbell at a distance
$d_{2}$ implies that the quadrupole tensor (Eq. \ref{eq:STF_quad})
has now:
\begin{equation}
\mathcal{Q}=\frac{m_{1}m_{2}}{m_{b}}d_{2}^{2}
\end{equation}
and the mismatch in the quadrupole tensor, with respect to the equal
mass dumbbell, is:
\begin{equation}
\Delta Q=\left(\frac{m_{b}}{4}-\frac{m_{1}m_{2}}{m_{b}}\right)d_{2}^{2}
\end{equation}
The difference in the potential will scale in a
similar way:
\begin{equation}
\Delta\Phi_{2}\sim\frac{G\left|\Delta Q\right|}{r^{3}}\propto G\left(\frac{m_{b}}{4}-\frac{m_{1}m_{2}}{m_{b}}\right)\frac{R_{x}^{2}}{r^{3}}\ell
\end{equation}
and for a test particle at a distance $r=\kappa R_{x}$, the relative
difference with respect to the monopole is
\begin{equation}
\frac{\left|\Delta\Phi_{2}\right|}{\left|\Phi_{0}\right|}\sim\left(\frac{1}{4}-\frac{m_{1}m_{2}}{m_{b}^{2}}\right)\frac{\ell}{\kappa^{2}}=\left(\frac{1}{4}-\frac{\eta}{(1+\eta)^{2}}\right)\frac{\ell}{\kappa^{2}}\label{eq:quadrupol_manon}
\end{equation}
with the same scaling for the accelerations. In the case of Quaoar,
for a mass fraction $\eta=0.1$, this gives $\sim1.4\times10^{-3}$
and $\sim5\times10^{-4}$ for the inner and outer rings, respectively.
The maximum difference arises when $\eta=0$, which
gives an approximate measure of the total quadrupole contribution
of Quaoar's ellipsoid with respect to the monopole. For completeness,
we obtain that at the octupole term, the relative difference
will be:
\begin{equation}
\frac{\left|\Delta\Phi_{3}\right|}{\left|\Phi_{0}\right|}\sim\frac{m_{1}m_{2}}{m_{b}^{3}}(m_{1}-m_{2})\frac{\ell^{3/2}}{\kappa^{3}}=\frac{\eta(1-\eta)}{(1+\eta)^{3}}\frac{\ell^{3/2}}{\kappa^{3}}\label{eq:octupol_manom}
\end{equation}

\subsection{The $J_{2}$ and $C_{22}$ connection and the topology of the SOR}\label{subsec:The-J2-and}

When dealing with rigid body dynamics, it is usual to write the multipolar
expansion of the potential in terms of certain coefficients, of which
the ones that appear at quadrupole order are $J_{2}$, associated
to the polar oblateness, and $C_{22},S_{22}$, associated to the equatorial
asymmetry \citep{1999ssd..book.....M}. These coefficients are nothing
but a repackaging of the STF quadrupole tensor, and their relation
can be found by comparing the quadrupolar potential in terms of the
STF tensor $Q_{ij}$:
\begin{equation}
\Phi_{2}(\vec{r})=-\frac{G}{2r^{5}}Q_{ij}r_{i}r_{j},\qquad i,j=x,y,z
\end{equation}
with the quadrupolar potential in terms of the spherical harmonics:
\begin{equation}
\Phi_{2}(r,\theta,\phi)=-\frac{Gm_{b}}{r^{3}}R^{2}\left[J_{2}\left(\frac{1}{2}-\frac{3}{2}\cos^{2}\theta\right)+\left(C_{22}\cos2\phi+S_{22}\sin2\phi\right)6\sin^{2}\theta\right]\label{eq:sphrical_harm}
\end{equation}
where $R$ represents the reference radius of the body. In the case
of the binary dumbbell, which is not an extended body, this reference
radius becomes arbitrary. It could be chosen, for example, to be the
mean radius of the ellipsoid that the dumbbell intends to mimic. Writing
the components $r_{i},r_{j}$ in spherical coordinates and using some
trigonometric identities, we obtain that:
\begin{equation}
J_{2}=-\frac{Q_{zz}}{2m_{b}R^{2}},\qquad C_{22}=\frac{Q_{xx}-Q_{yy}}{24m_{b}R^{2}},\qquad S_{22}=\frac{Q_{xy}}{12m_{b}R^{2}}
\end{equation}
For the equal mass binary dumbbell, in the body fixed frame, we get
\begin{equation}
J_{2}^{(b)}=\frac{a_{2}^{2}}{8R^{2}},\qquad C_{22}^{(b)}=\frac{a_{2}^{2}}{32R^{2}},\qquad S_{22}^{(b)}=0.
\end{equation}
In the inertial frame, and taking into account Eq. (\ref{eq:STF_quad}),
we get
\begin{equation}
J_{2}=J_{2}^{(b)},\qquad C_{22}=C_{22}^{(b)}\cos2n_{2}t,\qquad S_{22}=C_{22}^{(b)}\sin2n_{2}t\label{eq:j2c22t}
\end{equation}
Similar relations arise for the prolate ellipsoid and the mass anomaly
model. In all cases, the potential at quadrupole order is composed
of a $J_{2}$ fixed term plus an oscillating $C_{22},S_{22}$ term
with frequency $2n_{2}\equiv2\nu_{2}$. As expected, if we average
the potential over a period of the binary, $\overline{C}_{22}=\overline{S}_{22}=0$,
and the potential becomes a pure $J_{2}$ term, i.e. a rotating binary
behaves, on average, as an oblate spheroid.

Equations (\ref{eq:sphrical_harm}) and (\ref{eq:j2c22t}) provide
an insight into the expected topology of a SOR. The potential in the
inertial frame, at quadrupole order, contains an harmonic of the form
\begin{equation}
A_{2}\left(\frac{a}{r}\right)^{3}\cos(2\phi-2\lambda_{2})
\end{equation}
(cf. Eq. \ref{eq:rfsor}) where $\phi=f+\varpi$ represents the true
longitude of the test particle (we restrict to the co-planar motion
for simplicity), and
\[
A_{2}=-G\mu_{2}\frac{3}{4}\frac{a_{2}^{2}}{a^{3}}\equiv-\frac{Gm_{b}}{a^{3}}R^{2}6C_{22}^{(b)}<0.
\]
Expanding the true anomaly $f$ in terms of the mean anomaly $M$,
we have
\begin{equation}
\left(\frac{a}{r}\right)^{3}\exp\mathrm{i}2f=\sum_{j}\mathcal{X}_{j}^{-3,2}(e)\exp\mathrm{i}jM
\end{equation}
where $\mathcal{X}_{j}^{m,n}(e)$ are the Hansen coefficients \citep{1990CeMDA..49..209B}.
This makes to appear harmonics of the form
\begin{align}
\left(\frac{a}{r}\right)^{3}\cos(2f+2\varpi-2\lambda_{2}) & =\sum_{j}\mathcal{X}_{j}^{-3,2}(e)\cos(jM+2\varpi-2\lambda_{2})\nonumber \\
 & =\sum_{j}\mathcal{X}_{j}^{-3,2}(e)\cos(j\lambda-(j-2)\varpi-2\lambda_{2})\nonumber \\
 & =\sum_{j}\mathcal{X}_{j}^{-3,2}(e)\cos\left(j\frac{1}{k}\sigma+\left(j\frac{l}{k}-2\right)\lambda_{2}-\left(j\frac{l}{k}-2\right)\varpi\right)
\end{align}
where we replace $\lambda=(\sigma+l\lambda_{2}+(k-l)\varpi)\,/\,k$
from the definition of the resonant angle for a $k$:$l$ resonance
(Eq. \ref{eq:reso_angle}). After averaging over $\lambda_{2}$ (Eq.
\ref{eq:averager}), only the harmonics with $j=2k\,/\,l$ will survive
and we could write
\begin{align}
A_{2}\left(\frac{a}{r}\right)^{3}\cos(2\phi-2\lambda_{2}) & =\mathcal{A}_{s}^{k,l}(e)\cos s\sigma\label{eq:s-harmonic}
\end{align}
with $s=j/k=2/l$, and $\mathcal{A}_{s}^{k,l}=A_{2}\mathcal{X}_{sk}^{-3,sl}$.
Since $\mathcal{A}_{s}^{k,l}<0$ over a wide range of eccentricities,
then according to Eqs. (\ref{eq:equilib-1})-(\ref{eq:equilib}),
stable equilibrium will hold at $s\sigma_{0}=\pi,3\pi,\ldots$. Therefore,
if $l=2$, the only possible harmonic at quadrupole order is $s=1$,
implying that for resonances like 3:2, 5:2, 7:2, etc., there will
be only one libration equilibrium at $\sigma_{0}=180^{\circ}$. On
the other hand, if $l=1$, the only possible harmonic at quadrupole
order is $s=2$, thus for resonances like 2:1, 3:1, 4:1, etc. there
will be two libration equilibria at $\sigma_{0}=90^{\circ},270^{\circ}$.
An alternative way of interpreting this result is to redefine the
resonant angle as $\sigma^{\prime}=s\sigma$, and realize that the
2:1, 3:1, 4:1, etc. resonances are in fact of the form 4:2, 6:2, 8:2,
etc., i.e. resonances of order $2(k-l)$ even. On the other hand,
the resonances 3:2, 5:2, 7:2, etc., are always of order $k-l$ odd.
Here, we will maintain the base $k$:$l$ nomenclature related to
the angle $\sigma$, but we will consider the true order of the corresponding
resonance.

The location of the stable equilibrium points represents an important
difference with respect to the classical SOR problem arising in the
context of tidal torques, where the libration center is at $\sigma_{0}=0^{\circ},180^{\circ}$
for the 1:1 resonance, or at $\sigma_{0}=0^{\circ}$ for the 3:2 resonance
(cf. Sect. \ref{sec:Introduction}). In the classical SOR problem,
where the extended body feels a torque in its own quadrupole, the
resonant condition is defined by $kn\approx2\dot{\psi}$ and the averaging
is not performed over the rotational phase $\psi$, but over all the
non resonant harmonics $(j\neq k)$. Nevertheless,
the structure of the disturbing function at quadrupole order is still
of the form $\mathcal{B}_{s}^{k,l}(e)\cos s\sigma$, with $\mathcal{B}_{s}^{k,l}<0$
(cf. Eq. \ref{eq:rcsor}), but now the unperturbed Hamiltonian is
$\mathcal{F}\propto p_{\psi}^{2}/2>0$ and $\mathcal{F}_{p_{\psi}p_{\psi}}>0$,
thus equilibrium holds at $s\sigma_{0}=0,2\pi,\ldots$. This subtle
change of sign in the second derivative of $\mathcal{F}$ is what
differentiates the SORs around a rotating irregular body from the
classical conservative model of spin-orbit resonances \citep{2007PSS...55..889C}.

Coming back to Eq. (\ref{eq:s-harmonic}), for SORs
of degree $l\geq3$, there are no possible values of $s$ to represent
the harmonic at quadrupole order, so it is necessary to consider higher
even-order multipoles, getting that, in general, the averaged disturbing
function for a $k$:$l$ resonance becomes represented by a series
\begin{equation}
\mathcal{R}=\sum_{s}\mathcal{A}_{s}^{k,l}(e)\cos s\sigma\label{eq:harmonic}
\end{equation}
where
\begin{equation}
\mathcal{A}_{s}^{k,l}=\sum_{i\,\geq\,\max(2,sl)}\frac{1+(-1)^{i-sl}}{2}A_{i}\mathcal{X}_{sk}^{-(i+1),sl}\label{eq:coeff}
\end{equation}
and $i=2,4,6,\ldots$ are the orders of the multipoles. For $l=4$,
for example, the leading term ($s=1$) only appears at the hexadecapole
($i=4$), while for $l=3$, the leading term ($s=2$) only appears
at the tetrahexacontapole ($i=6$). In general, the value of $\mathcal{A}_{s}^{k,l}$
may switch from negative to positive depending on $e$, which will
make the libration centers to shift to $\sigma_{0}=0^{\circ}$ if
$l$ is even, and to $\sigma_{0}=0^{\circ},180^{\circ}$ if $l$ is
odd. The above result is valid for any potential with azimuthal $\pi$-symmetry,
including prolate and triaxial ellipsoids (provided that the triaxial
ellipsoid rotates around its minor axis); just changing $\lambda_{2}$
by the phase of the major axis of the body. 

The location of the SOR libration centers and the resonance widths
display a different behavior in the mass anomaly problem, or any problem
where the octupole and other odd-order multipoles are relevant, as
we will see in Sect. \ref{subsec:Mass-anomaly-system}.

\subsection{The SOR widths}

The topology of the SOR is not only determined by the location of
the equilibrium points, but also by the width of the resonance. The
disturbing function, Eq. (\ref{eq:distrubing}), scales at lowest
(quadrupole) order as 
\begin{equation}
\mathcal{R}\propto G\frac{m_{1}m_{2}}{m_{b}}\frac{a_{2}^{2}}{a^{3}}=\frac{m_{1}m_{2}}{m_{b}^{2}}a_{2}^{2}n^{2},\label{eq:r-leading}
\end{equation}
and according to Eq. (\ref{eq:widths}), the width of the resonance
scales as $\sqrt{\Delta\mathcal{R}}/n$. Thus, if the binary separation
$a_{2}$ is kept fixed while varying the masses, the resonance width
will scale as 
\begin{equation}
\Delta a\propto\frac{\sqrt{m_{1}m_{2}}}{m_{b}}=\frac{\sqrt{\eta}}{1+\eta},\label{eq:width-scale}
\end{equation}
being maximum when $\eta=1$ and shrinking as $\eta$ decreases. If
the mass ratio is kept fixed while varying the binary separation,
the resonance width will scale linearly with $a_{2}$. On
the other hand, the resonance width is independent of $n$, and thus
it will be independent of the orbital frequency of the binary $n_{2}$.
Therefore, depending on the values of $\eta,a_{2}$, SORs may overlap
and this might be relevant to destabilize the space around the binary
and, by direct analogy, around any irregular shaped body. On the other
hand, for $\eta,a_{2}$ fixed, we should expect that the higher the
degree $l$ of the SOR, the smaller the strength (and so the width)
of the resonance, because it is driven by higher order multipoles.

\subsection{Parameters of the Quaoar system}

In our applications of the circumbinary model to the Quaoar system,
we consider the parameters presented in Tables \ref{tab:Physical-parameters-of}
and \ref{tab:Initial-orbital-elements}. The listed $R_{y}$ radius
has been computed as the average of the true reported radii $R_{y},R_{z}$.
The mass of Quaoar is taken as the total mass of the binary, $m_{b}$.
Distances are given in units of Quaoar's mean radius. The orbital
inclinations are referred to the ecliptic, and the inclination of
the binary corresponds to the axis tilt of Quaoar.

\begin{table}

\caption{Physical parameters of the Quaoar system.}\label{tab:Physical-parameters-of}

\centering{}\setlength{\tabcolsep}{0.8em}%
\begin{tabular}{lr@{\extracolsep{0pt}.}lr@{\extracolsep{0pt}.}lcccc}
\hline 
 & \multicolumn{2}{c}{Mass $(\times10^{-4}M_{\oplus})$} & \multicolumn{2}{c}{$\overline{R}$ (km)} & $R_{x}$ (km) & $R_{y}$ (km) & $T_{\mathrm{rot}}$ (h) & Reference\tabularnewline
\hline 
Quaoar & 2&0294 & 549&0 & 583.3 & 532.65 & 17.6788 & \citet{Margoti2024}\tabularnewline
Weywot & 0&00729 & 82&5 & -- & -- & -- & \citet{2025PSJ.....6..285P}\tabularnewline
Second moon & 0&00007 & 15&0 & -- & -- & -- & \citet{2025ApJ...993L..38P}\tabularnewline
\hline 
\end{tabular}
\end{table}

\begin{table}
\caption{Initial orbital elements of the Quaoar system.}\label{tab:Initial-orbital-elements}

\begin{centering}
\setlength{\tabcolsep}{0.9em}%
\begin{tabular}{lr@{\extracolsep{0pt}.}lr@{\extracolsep{0pt}.}lr@{\extracolsep{0pt}.}lr@{\extracolsep{0pt}.}lr@{\extracolsep{0pt}.}lr@{\extracolsep{0pt}.}lr@{\extracolsep{0pt}.}l}
\hline 
 & \multicolumn{2}{c}{$a$ ($R_{Q}$)} & \multicolumn{2}{c}{$e$} & \multicolumn{2}{c}{$i$ $(^{\circ})^{\,3}$} & \multicolumn{2}{c}{$\lambda$ $(^{\circ})$} & \multicolumn{2}{c}{$\varpi$ $(^{\circ})$} & \multicolumn{2}{c}{$\Omega$ $(^{\circ})$} & \multicolumn{2}{c}{Reference}\tabularnewline
\hline 
Binary secondary & 0&387$^{1}$ & 0&0 & 12&6 & 270&0$^{2}$ & 0&0 & 0&0 & \multicolumn{2}{c}{\citet{2025PSJ.....6..146P}}\tabularnewline
Weywot & 24&279 & 0&011 & 13&6 & 169&3 & 90&3 & 353&3 & \multicolumn{2}{c}{\citet{2025PSJ.....6..285P}}\tabularnewline
Second moon & 10&634 & 0&0 & 12&6 & 0&0 & 0&0 & 0&0 & \multicolumn{2}{c}{\citet{2025ApJ...993L..38P}}\tabularnewline
Inner ring particle & 4&590 & 0&0 & 12&6 & 0&0 & 0&0 & 0&0 & \multicolumn{2}{c}{\citet{2023AA...673L...4P}}\tabularnewline
Outer ring particle & 7&390$^{4}$ & 0&0 & 18&4 & 0&0 & 0&0 & 0&0 & \multicolumn{2}{c}{\citet{2026ApJ...999L..39B}}\tabularnewline
\hline 
\end{tabular}
\par\end{centering}
\begin{centering}
$^{1}$ From Eq. (\ref{eq:calib_dist}).
\par\end{centering}
\begin{centering}
$^{2}$ From Eq. (\ref{eq:lambda}).
\par\end{centering}
\begin{centering}
$^{3}$ Inclinations are referred to the ecliptic.
\par\end{centering}
\centering{}$^{4}$ Maximum estimated variation $\pm0.27$.
\end{table}

\section{Results}\label{sec:Results}

\subsection{Equal mass binary system}

We start by studying the equal mass binary model and mapping the SORs
that occur in the Quaoar system. We explore the range between $4\,R_{q}$
and $9\,R_{Q}$, that spans the locations of the inner and outer rings.
Unless specified differently, the parameters used in the plots are
those of Tables \ref{tab:Physical-parameters-of} and \ref{tab:Initial-orbital-elements}.

\begin{figure}

\caption{The widths of the main SORs and MMRs in the Quaoar system computed
with the semianalytical model. The different colors indicate the true
order $q$ of the resonance: red $q=1$, green $q=2$, blue $q=3$,
gray $q=4$, cyan $q=5,$yellow $q=6$, magenta $q=7$. The dashed
curves represent the collision/instability limits. All resonances
were computed in the planar model. The locations and widths of the
rings Q1R and Q2R are indicated at the bottom.}\label{fig:widths-i0}

\begin{centering}
\includegraphics[width=0.5\textwidth]{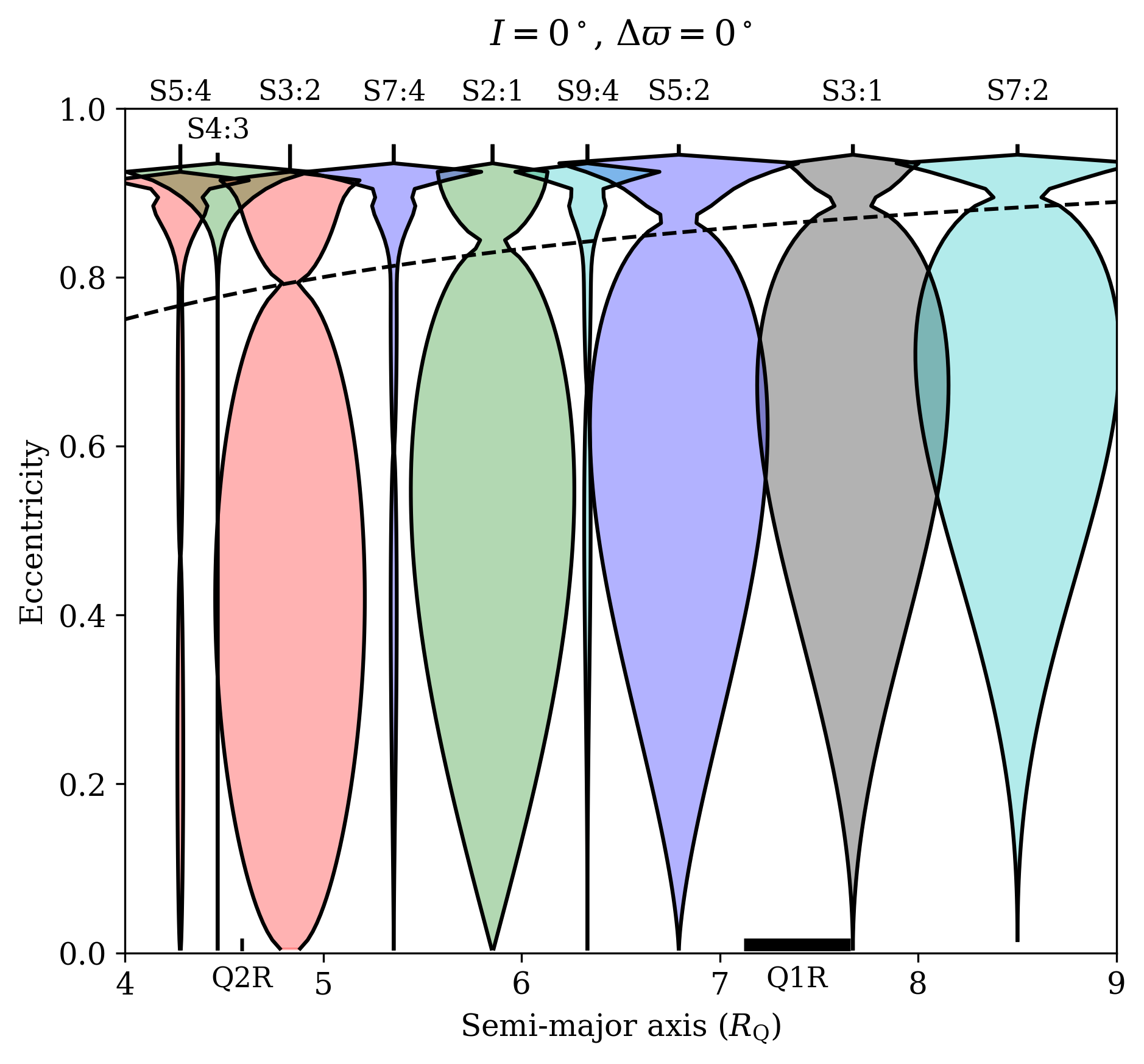}\includegraphics[width=0.5\textwidth]{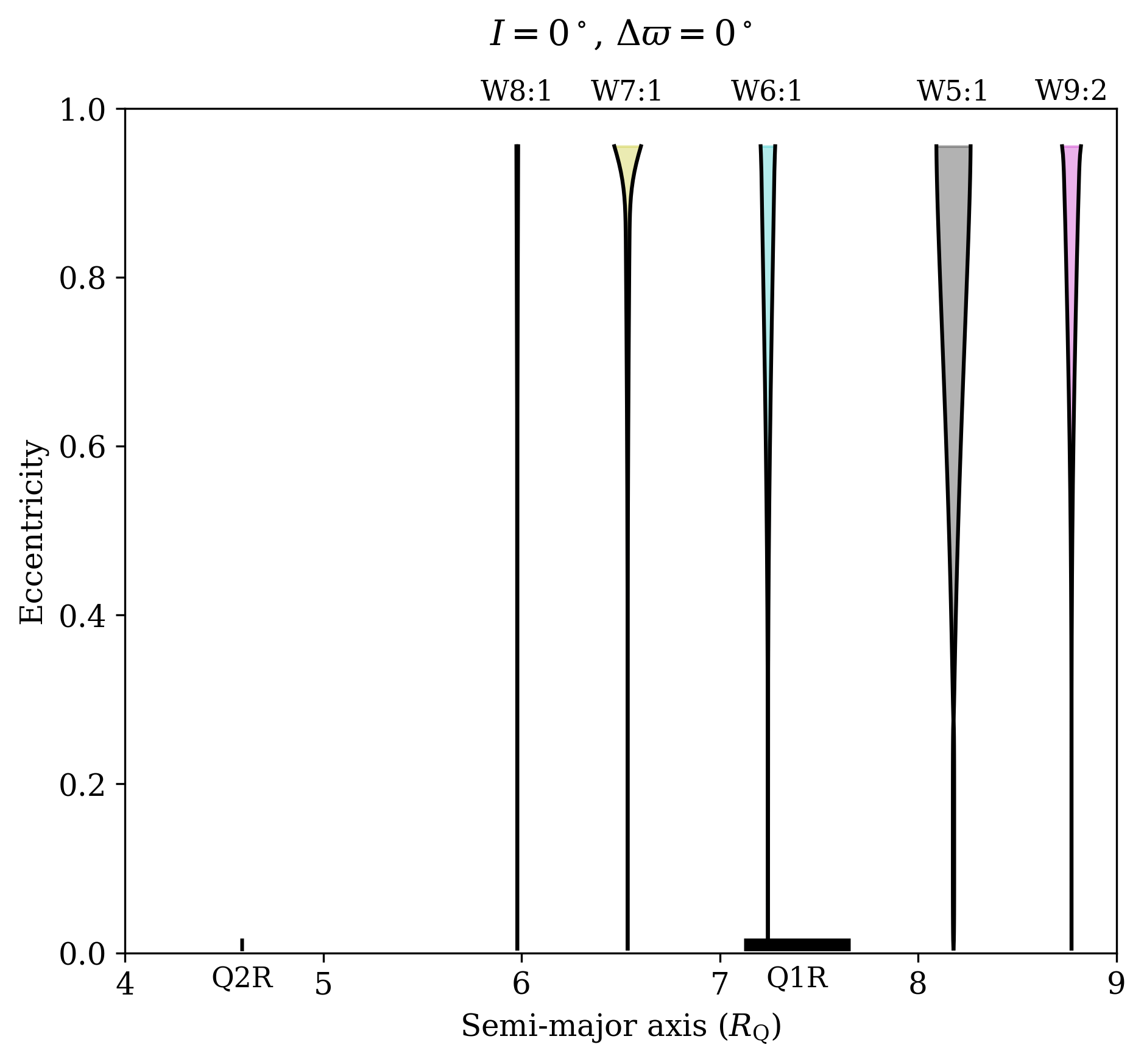}
\par\end{centering}
\centering{}\includegraphics[width=0.5\textwidth]{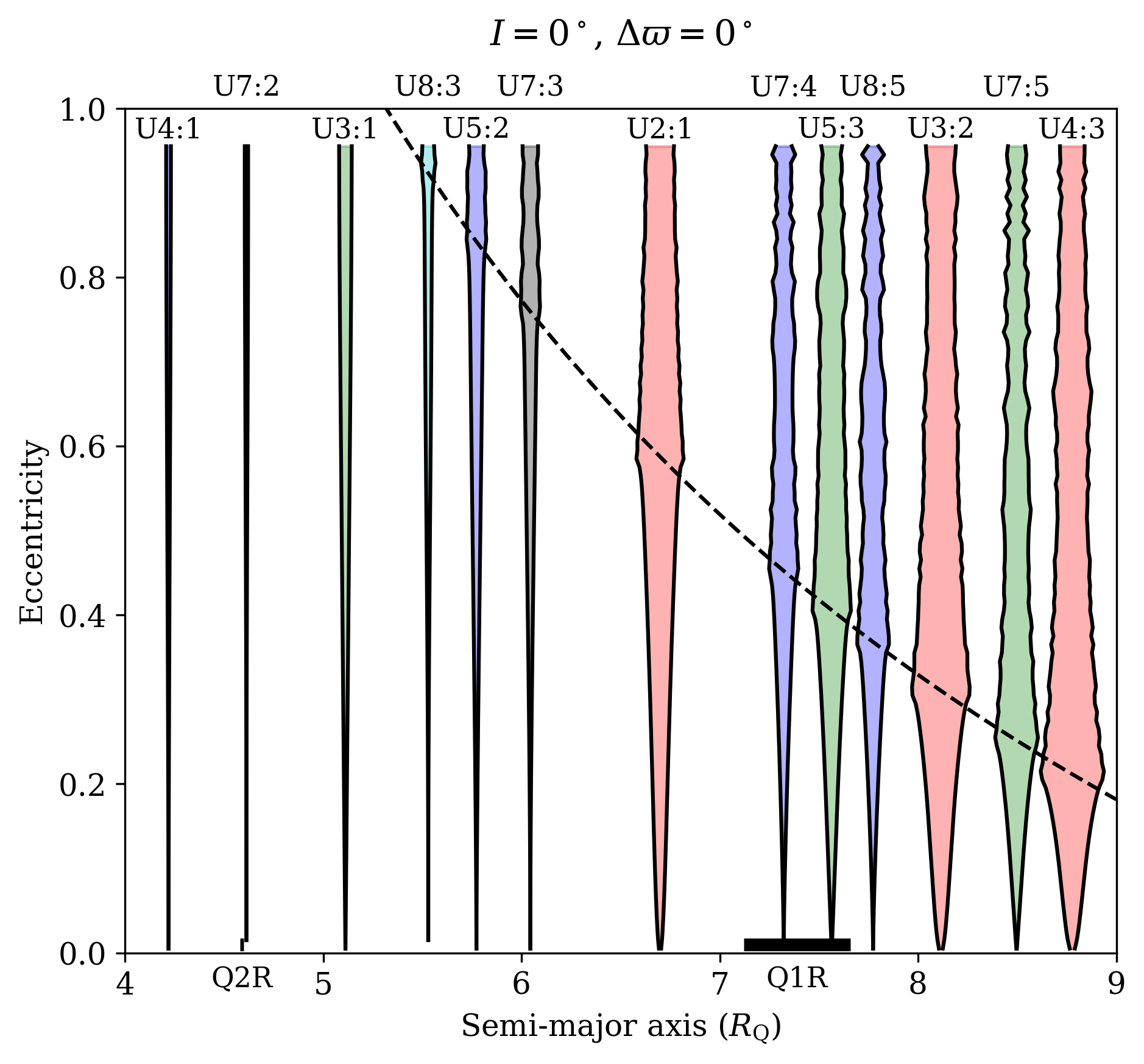}
\end{figure}

Figure \ref{fig:widths-i0} shows the web of SORs and MMRs around
the Quaoar system. Panel (a) shows the SORs, panel (b) shows the MMRs
with Weywot, and panel (c) shows the MMRs with the second unconfirmed
moon. The widths have been computed with the semianalytical model,
and the colors identify the true order $q$ of the resonance. For
example, the 2:$1\equiv4$:2 SOR is actually an order 2 resonance.
The approximate location and extension of the two rings, Q1R and Q2R,
is indicated in the horizontal axis. The computations have been done
in the co-planar case ($I=0^{\circ}$ for all the bodies) . We note
that the low degree SORs are significantly wider than the MMRs and
dominate the spanned interval. At very high eccentricities, the resonance
overlap is expected to create a region of chaotic evolution. On the
other hand, the higher degree SORs ($l=3,4$) appear very narrow,
as expected. It is especially noteworthy that the 4:3 SOR is narrower
than the 5:4, 7:4 and 9:4 SORs, which seems counterintuitive. But
we must recall that the 4:3 resonance is driven by a multipole of
higher order than the one that drives the 5:4, 7:4, etc. 

The SORs panel also displays the stability limit (dashed line) predicted
by \citet{1999AJ....117..621H} for circumbinary orbits, which in
our model is given by:
\begin{equation}
r_{\mathrm{crit}}=a_{2}\left[1.6+4.12\frac{m_{2}}{m_{b}}-5.09\left(\frac{m_{2}}{m_{b}}\right)^{2}\right]\approx1\,R_{Q}\label{eq:holman}
\end{equation}

Concerning the MMRs, those with Weywot are extremely thin and sparse.
The resonances with the unconfirmed moon are wider and denser, in
spite of this body have one hundredth of the mass of Weywot. The U5:3
and U7:4 MMRs overlap with the location of the outer ring, but as
we will see later, they do not have any relevant effect in the dynamics.
The dashed line in panel (c) corresponds to the collision with the
unconfirmed moon, defined by $a(1+e)=a_{\mathrm{U}}$.

\begin{figure}
\caption{The widths of the main SORs in the Quaoar system, for inclined test
particles, computed with the semianalytical model. }\label{fig:widths-i180}

\centering{}\includegraphics[width=0.5\textwidth]{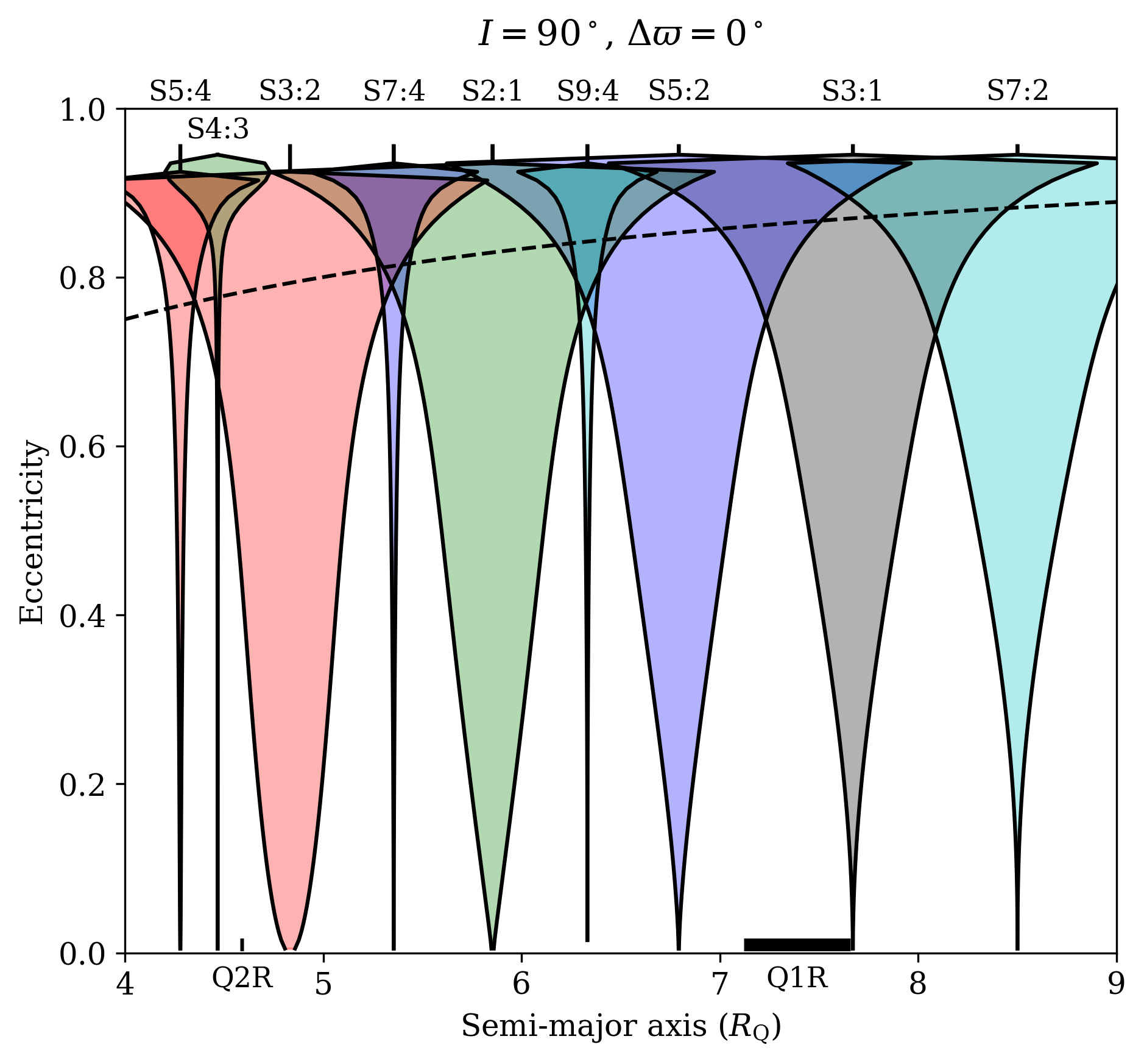}\includegraphics[width=0.5\textwidth]{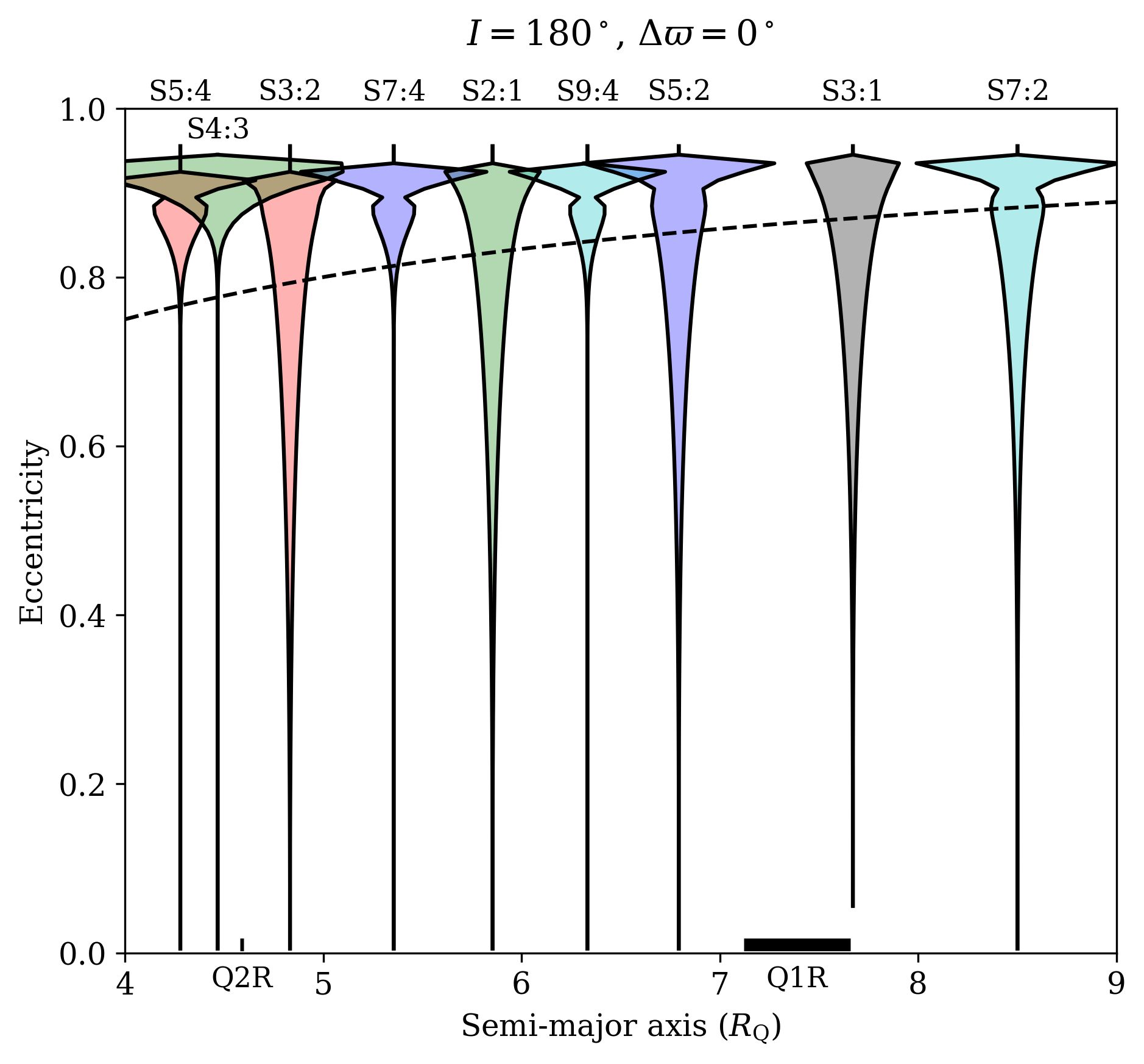}
\end{figure}

If we change the inclination of the test particle, with respect to
Quaoar's equator, we find that the SORs start to shrink. This is shown
in Fig. \ref{fig:widths-i180} for a polar orbit $(I=90^{\circ})$
and a pure retrograde orbit $(I=180^{\circ})$. This means that retrograde
orbits are much less affected by the SORs, and thus are expected to
be more stable in the long term. The stability of
retrograde motions could be relevant for the dynamics of the rings,
especially for the inner ring, as we will see in Sect. \ref{subsec:Analysis-of-specific}.

\begin{figure}
\caption{Topology of the 3:1 and 3:2 SORs in the Quaoar system, computed with
the semianalytical model. The level curves correspond to a test particle
with $e=0.25$ and $I=0^{\circ}$. It is worth noting that the librations
in the 3:1 SOR are symmetric. }\label{fig:topol}

\centering{}\includegraphics[width=0.5\textwidth]{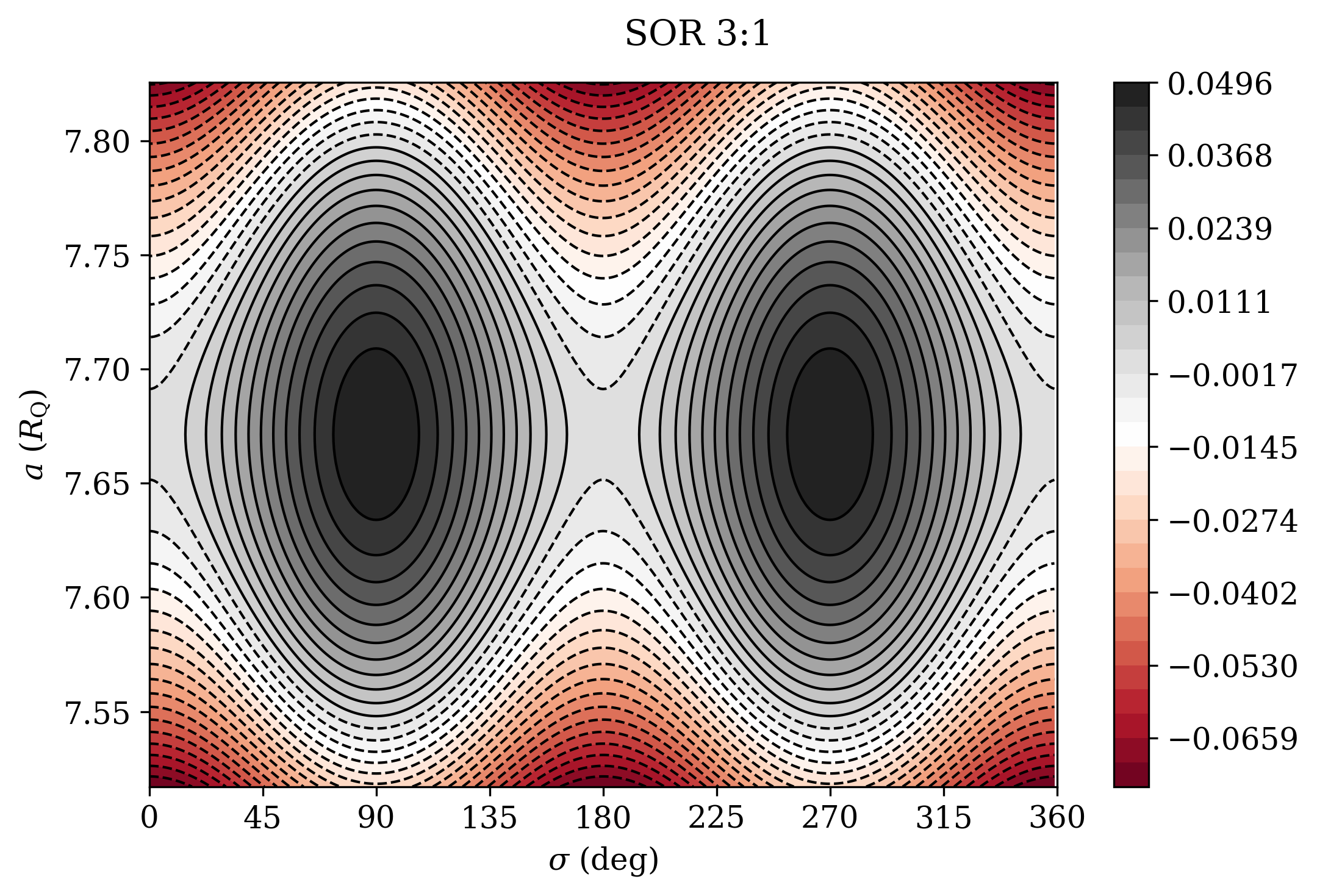}\includegraphics[width=0.5\textwidth]{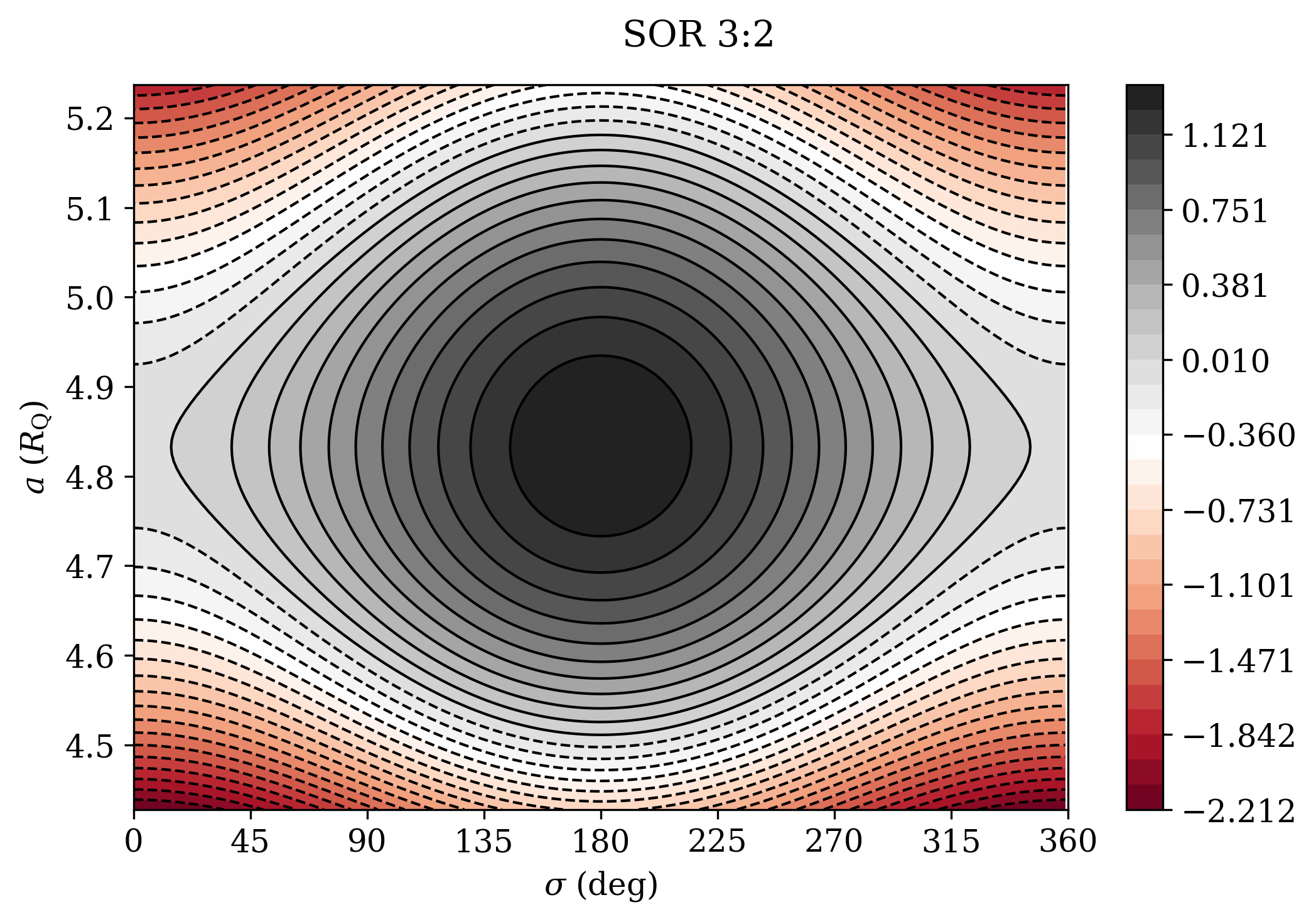}
\end{figure}

The application of the semianalytical model produces the phase space
topology that is exemplified in Fig. \ref{fig:topol}. These level
curves of the Hamiltonian at the 3:1 and 3:2 SORs have been obtained
for a test particle with $e=0.25$ and $I=0^{\circ}$. As expected,
in the 3:1 SOR ($l$ odd), the libration centers are at $\sigma_{0}=\pm90^{\circ}$,
while in the 3:2 SOR ($l$ even), the libration occurs around $\sigma_{0}=180^{\circ}$.
The result for the 3:1 SOR is consistent with the results of Gianuzzi
et al. (2026), when their two boulders model is configured in opposition.
For resonances with $l=1,2$, this pattern is almost independent of
both the eccentricity and the inclination of the test particle. The
particle tries to maintain a phase-locked orbit with the rotating
pattern of the potential, such as to cancel all the gravitational
torques over an orbit. In particular, when the particle is at pericenter
($\lambda=\varpi$), we get that
\[
\lambda-\lambda_{2}=\pm\frac{\pi}{2l}\quad(l\text{ odd}),\qquad\qquad\lambda-\lambda_{2}=\frac{\pi}{l}\quad(l\text{ even})
\]
and for resonances with $l=1,2$ this implies that the particle is
always perpendicular to the binary axis at its maximum approach. Only
at low inclinations ($I\lesssim20^{\circ}$) and very high eccentricities
($e\gtrsim0.8$), above the collision curve, this pattern breaks and
the libration centers suddenly shift from $\pm90^{\circ}$ to $0^{\circ},180^{\circ}$
for $l=1$, and from $180^{\circ}$ to $0^{\circ}$ for $l=2$. 

For resonances with larger values of $l$, the location of the libration
points follows a similar pattern, but the aforementioned shift of
the points may occur at lower eccentricities, typically $e\gtrsim0.4$
for $l=3,4$. This is consistent with the discussion presented in
Sect. \ref{subsec:The-J2-and}. 

For orbits well outside the binary plane ($20^{\circ}\lesssim I\lesssim160^{\circ}$),
the location of the libration points do not change with eccentricity
at all, and are kept fixed at $\sigma_{0}=\pm90^{\circ}$ for $l$
odd, and at $\sigma_{0}=180^{\circ}$ for $l$ even.

In any case, the behavior of the SOR equilibria is significantly different
from the behavior observed in MMRs, where the orbital configuration
of the bodies may lead to the occurrence of bifurcations and asymmetric
librations. None of these phenomena are observed in the SORs around
an equal mass binary. 

\begin{figure}
\caption{Libration period at SOR equilibria in the Quaoar system, computed
with the semianalytical model. The vertical dashed line is the collision/instability
limit. The periods are significantly overestimated
in the gray regions.}\label{fig:periods}

\centering{}\includegraphics[width=0.5\textwidth]{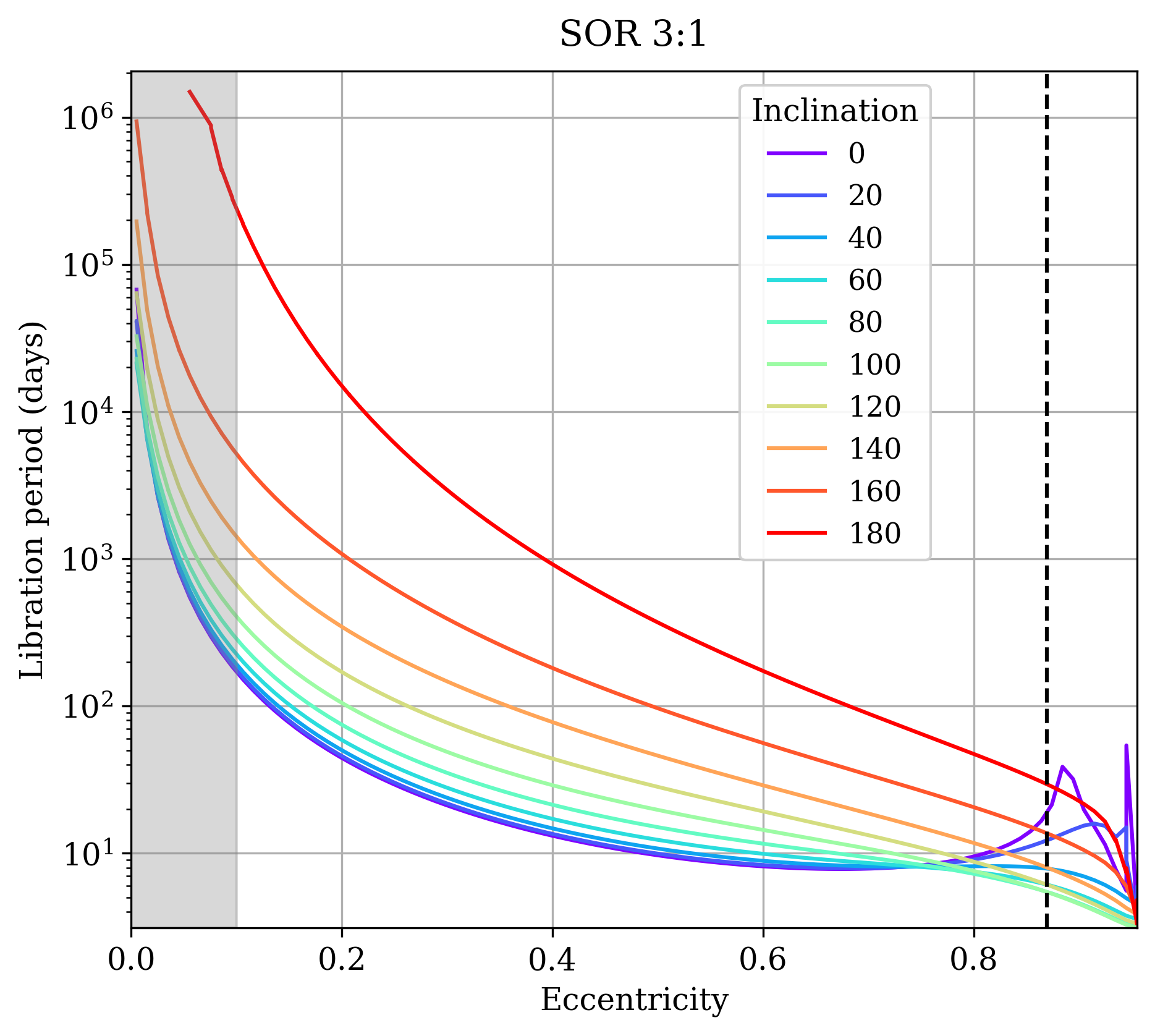}\includegraphics[width=0.5\textwidth]{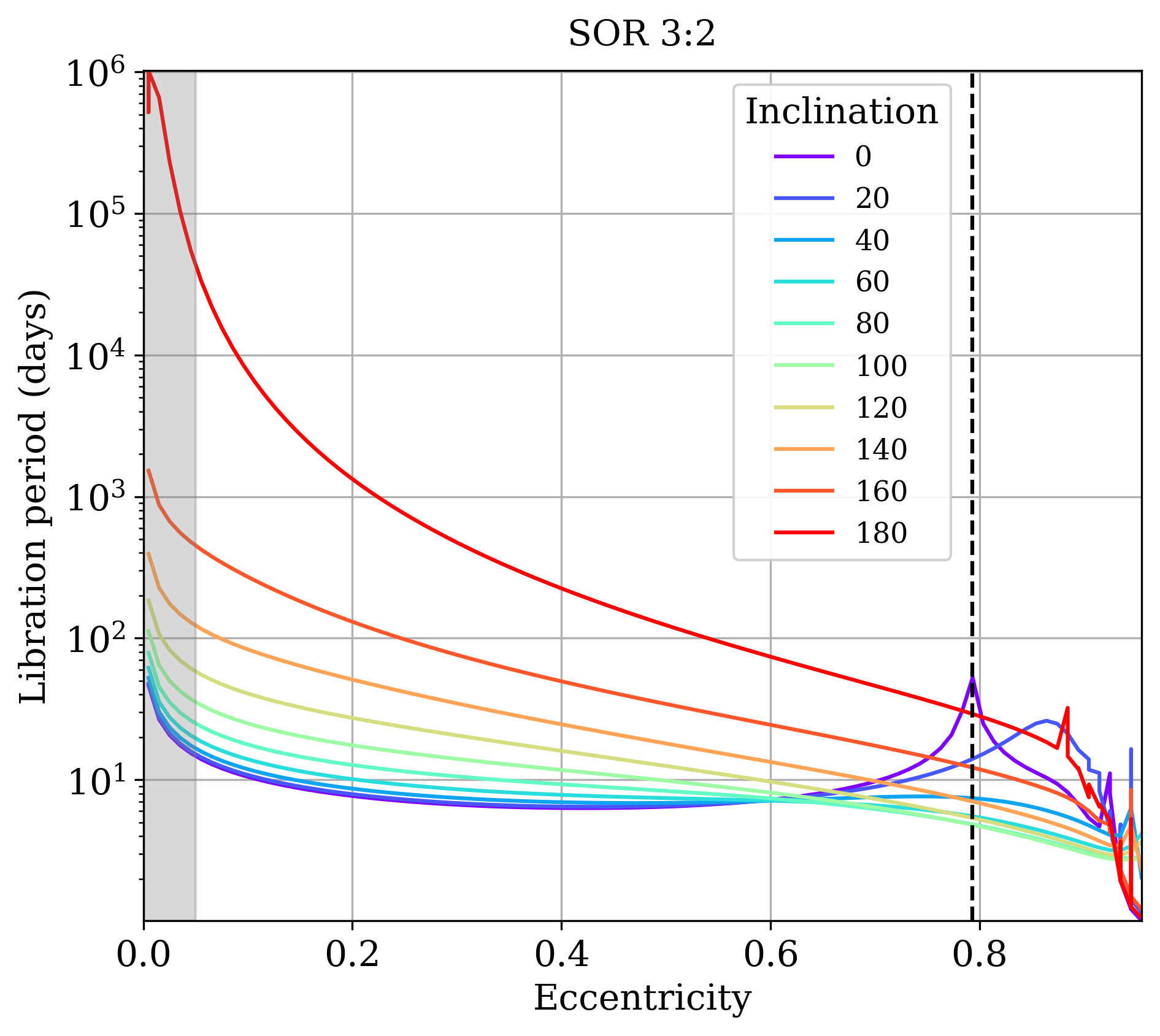}
\end{figure}

Figure \ref{fig:periods} exemplifies the behavior of the libration
period at the equilibrium points for the 3:1 and 3:2 SORs, in terms
of eccentricity and inclination. The libration period is always larger
than $\sim10$ times the orbital period of the binary. The libration
period increases with inclination, and the peaks that are observed
at very high eccentricities are likely caused by the collision condition
Eq. (\ref{eq:holman}). We shall note, however, that the period values
predicted for very low eccentricities ($e\lesssim0.05$-0.1) are significantly
overestimated, as we will show later, which is a limitation of the
semianalytical model.

\subsubsection{Dynamical maps}

\begin{figure}

\caption{Dynamical maps of the space around the Quaoar system considering different
configurations: (a) an equal mass binary; (b) the same binary plus
the satellite Weywot; and (c) the full system including the second
unconfirmed satellite. The white lines correspond to the resonance
borders, computed with the semianalytical model (for the main SORs
in panel (a) and some MMRs in panel (c). The approximate locations
of the rings are indicated below (white for the outer ring, and black
for the inner ring). For the outer ring we are considering the maximum
estimated width. The dashed line in panel (c) is the theoretical collision
curve with the unconfirmed satellite. Panel (d) is a zoom of panel
(c) with a more stringent color scale. The five magenta stars indicated
in panel (c) are specific orbits discussed in Sect. \ref{subsec:Analysis-of-specific}.
}\label{fig:dynamical-maps}

\begin{centering}
\includegraphics[width=0.49\textwidth]{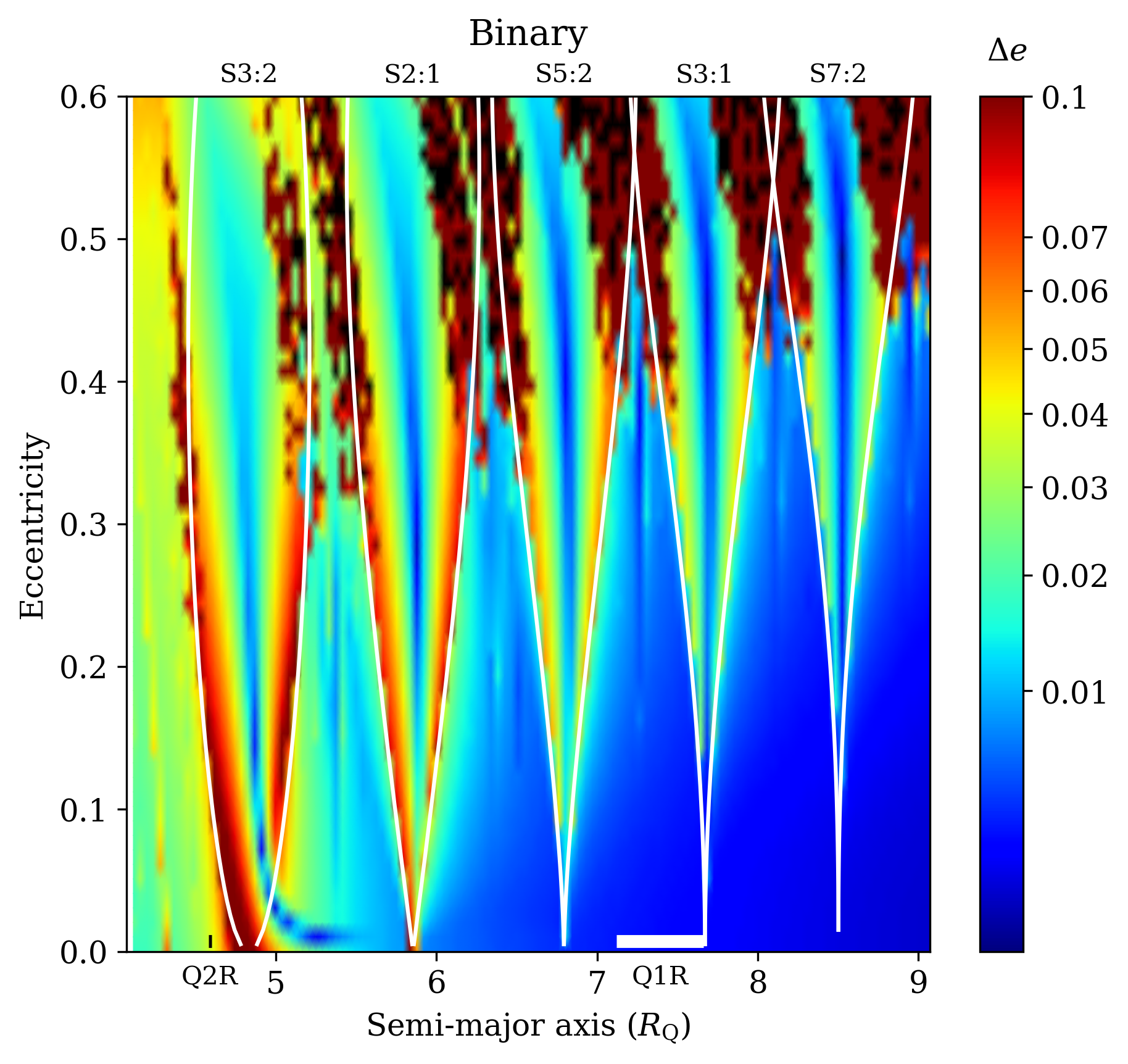}\includegraphics[width=0.49\textwidth]{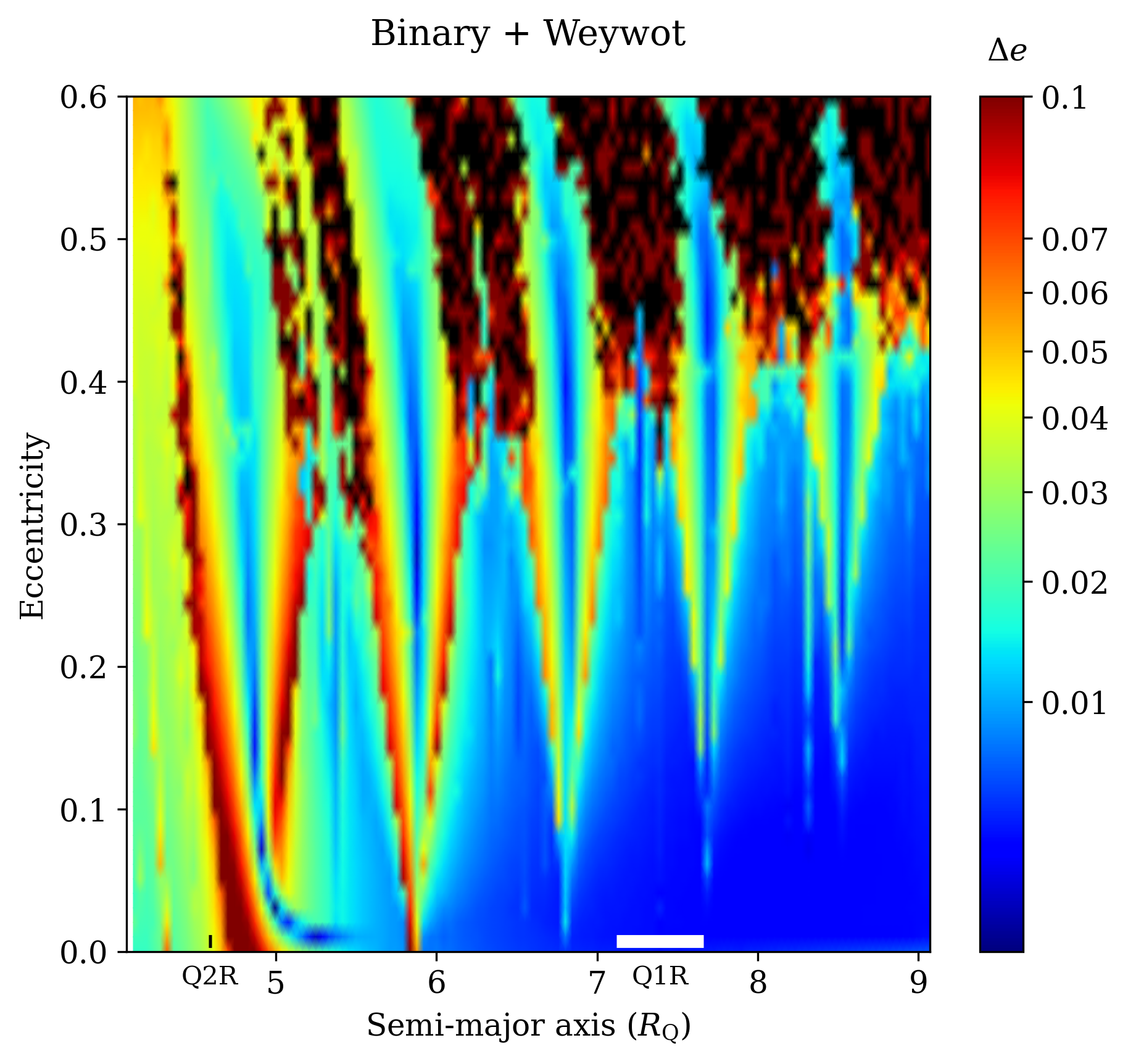}
\par\end{centering}
\centering{}\includegraphics[width=0.49\textwidth]{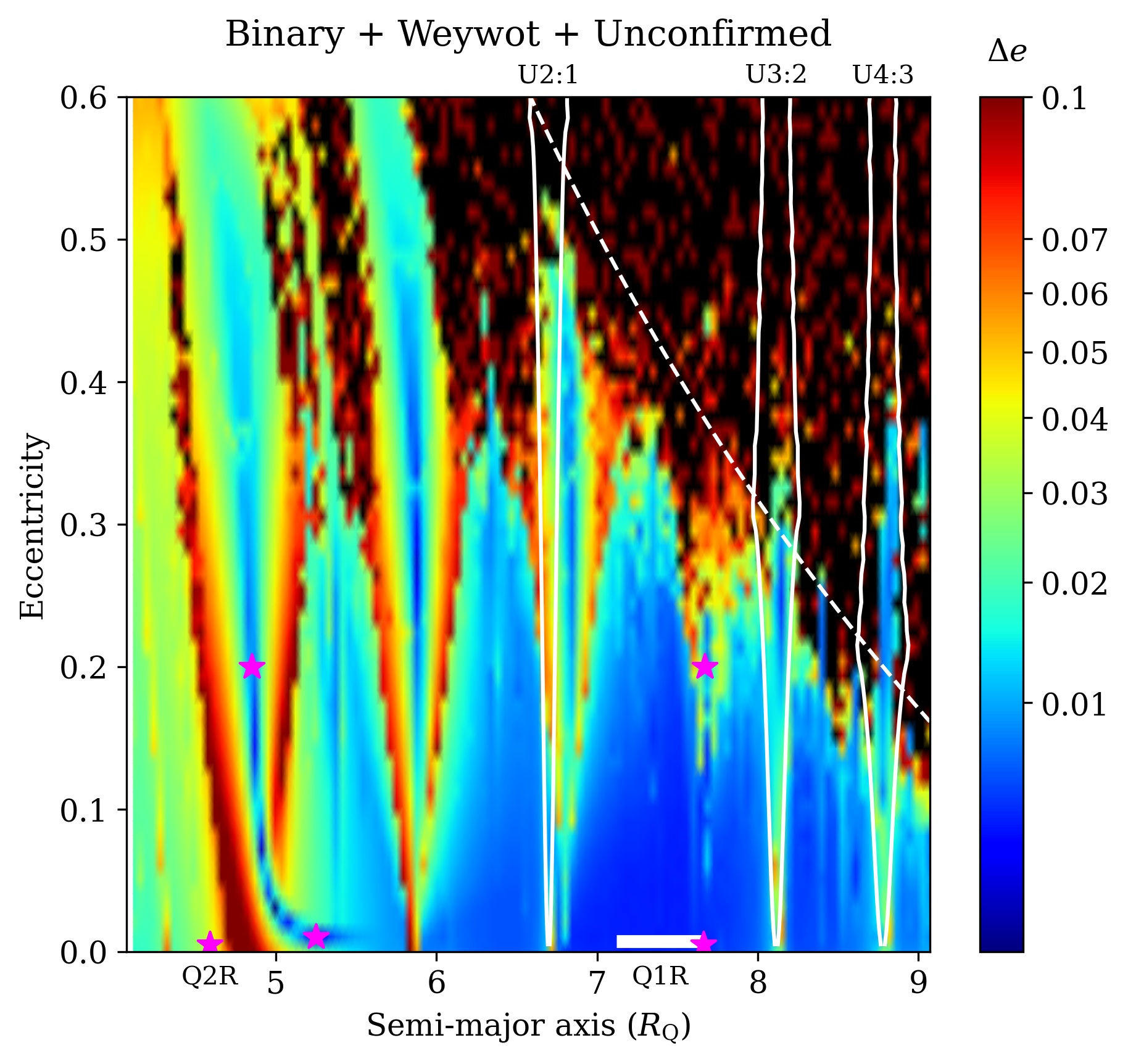}\includegraphics[width=0.49\textwidth]{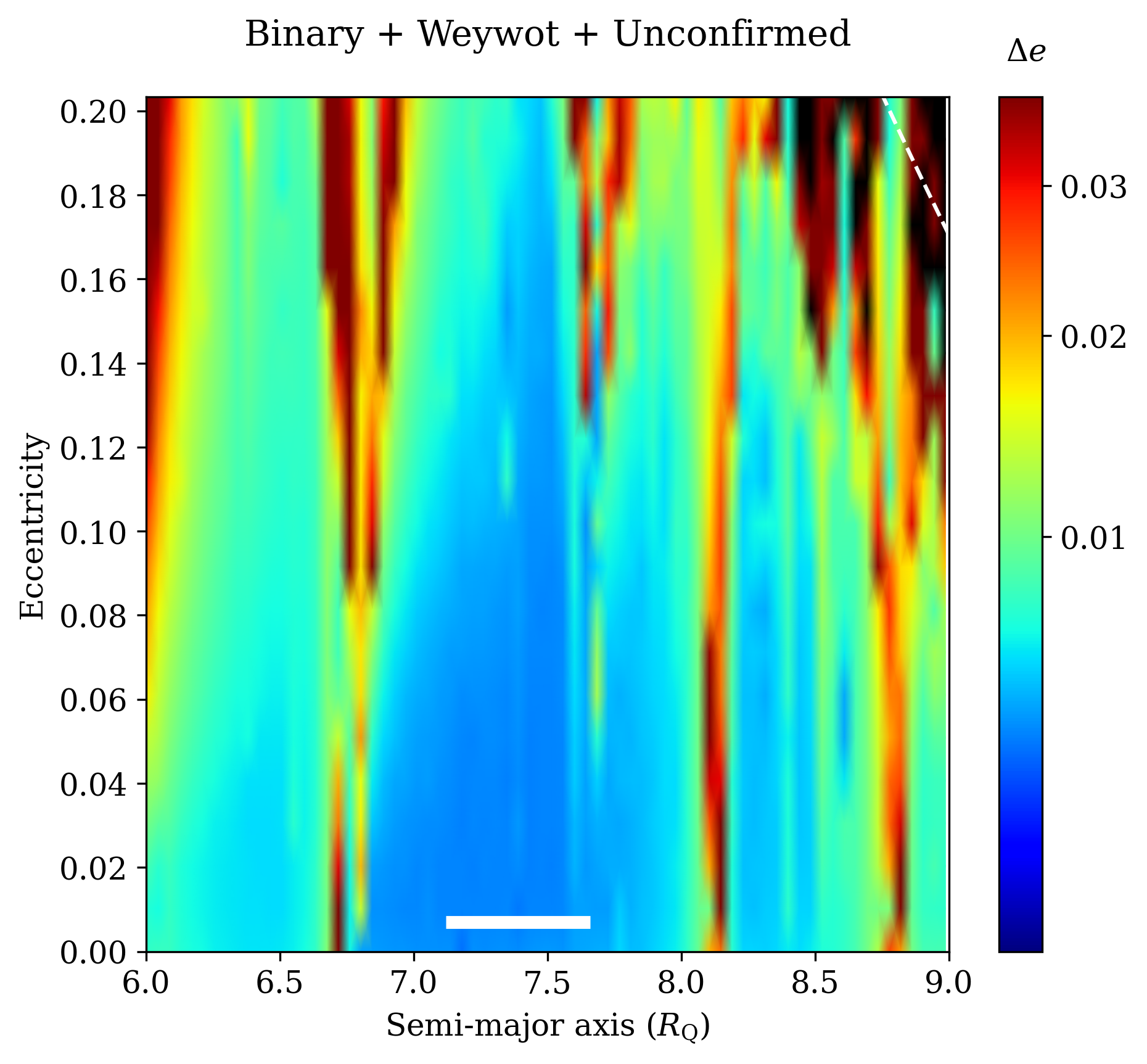}
\end{figure}

Applying the modified symplectic integrator to the Hamiltonian described
in Sect. \ref{subsec:Numerical-approach}, we produce a series of
dynamical maps of the Quaoar system. For these numerical simulations,
we used the actual parameters reported in Tables \ref{tab:Physical-parameters-of}
and \ref{tab:Initial-orbital-elements}, except for the inclination
of the test particle that has been assumed equal to that of the binary.
The maps were constructed for a grid of $120\times60$ initial conditions,
spanning the intervals $4\leq a\leq9\,R_{Q}$ and $0\leq e\leq0.6$.
The results are presented in Fig. \ref{fig:dynamical-maps}. The color
scale corresponds to the maximum variation in eccentricity, $\Delta e=\left|e(t)-e_{0}\right|_{\mathrm{max}}$,
with respect to the initial condition, computed over a time interval
of 5000 days. This quantity is easy to compute and it is a good proxy
for stability. A test particle in the grid is discarded from the simulation
whenever happens to have a close approach to any massive body (either
the binary secondary or a satellite) at less than 1.5 physical radii.
The integrator also checks if the particles get too close or too far
from the binary primary. The integration time step corresponds to
1/200th of the period of the binary. 

Panel (a) in Fig. \ref{fig:dynamical-maps} shows a system where only
the binary perturbation is taken into account. The white lines are
the borders or separatrices of the SORs computed with the semianalytical
model, the same shown in Fig. \ref{fig:widths-i0}. We verify that,
in general, the semianalytical widths fit very well to the numerical
predictions. As expected, the central part of the SORs are quite stable,
while some instability appears close to the separatrices. At very
high eccentricities, the overlap of the resonances generates a region
of larger instability. It is worth noting, however, that the color
scale only accounts for eccentricity variations up to 0.1, so the
actual variations in the whole space explored are indeed very small.
Only the grid regions indicated by a black color correspond to test
particles that were discarded due to close encounters with the binary. 

We note that the 3:2 SOR deviates significantly from the semianalytical
prediction for $e\lesssim0.06$. Indeed, the stable equilibrium of
the resonance shifts to semimajor axes larger than the resonant value
as the eccentricity decreases. This is the same phenomenon observed
in first-order MMRs, that is usually referred to as the law of structure
of the resonance. It is a well know fact that our semianalytical model
is unable to reproduce the law of structure of any resonance, therefore
the observed discrepancy. We also note that the 2:1 SOR does not display
a similar law of structure, which is expected since, as already mentioned,
this is actually a second order resonance. 

Panel (b) in Fig. \ref{fig:dynamical-maps} shows a system where now
the perturbation of Weywot has been included. It is evident that Weywot
does not affect the dynamics of the test particles at low eccentricities.
Only at very high eccentricities, the overlap of the SORs with the
MMRs contributes to slightly increase the region of instability. 

Panel (c) in Fig. \ref{fig:dynamical-maps} shows the case where both
Weywot and the unconfirmed satellite are included. This time the dynamical
effect is dramatic. Not only the overlap of the SORs and the MMRs
generates instability at low eccentricities, but there is a large
region at mid-to-high eccentricities that is depleted by close encounters
with the unconfirmed satellite. Some MMRs are highlighted, where the
resonant protection mechanism is acting to avoid close encounters
with the satellite above the collision curve. We must clarify, however,
that the depletion observed is based on a simple criterion of mutual
distance threshold between the particles and the satellite, and it
is possible that the test particles can actually survive to many close
encounters with such very low massive body. The five stars shown in
magenta in the plot correspond to specific initial conditions that
will be analyzed in detail in Sect. \ref{subsec:Analysis-of-specific}.

A relevant result for the dynamics of Quaoar's rings is presented
in panel (d) of Fig. \ref{fig:dynamical-maps}. This panel shows a
zoom of panel (c), with the color scale adjusted to highlight the
subtle changes in $\Delta e$. We can see that the outer ring is located
precisely in the most stable region of all the space around Quaoar.
This might be just a coincidence, but it is a curious fact. Moreover,
we can see that neither the 3:1 SOR, nor the 6:1 MMR with Weywot seems
to have any relevant role in the dynamics of the ring. MMRs with the
unconfirmed satellite do not have any effect either. In principle,
the total variation in eccentricity at the presumed location of the
ring is of the order of 0.006 only. The situation is considerably
different in the inner ring, where the maximum variation in eccentricity
may reach values of 0.04. This is discussed with more detail in the
next section.

\subsubsection{Analysis of specific orbits}\label{subsec:Analysis-of-specific}

Here, we present and discuss the time evolution of five specific orbits
around the Quaoar system, indicated by the magenta stars in Fig. \ref{fig:dynamical-maps}(c).
Two of these correspond to high eccentricity resonant initial conditions,
one to a low eccentricity resonant condition, and the remaining two
to particles located at the inner and outer ring, respectively.

\begin{figure}
\caption{Examples of the orbital evolution ($a,e,\varpi,\sigma$) of three
resonant particles. The particle in the low-$e$ 3:2 SOR is located
over the law of structure of the resonance and displays a paradoxical
libration.}\label{fig:Examples-of-the}

\begin{centering}
\includegraphics[width=0.49\textwidth]{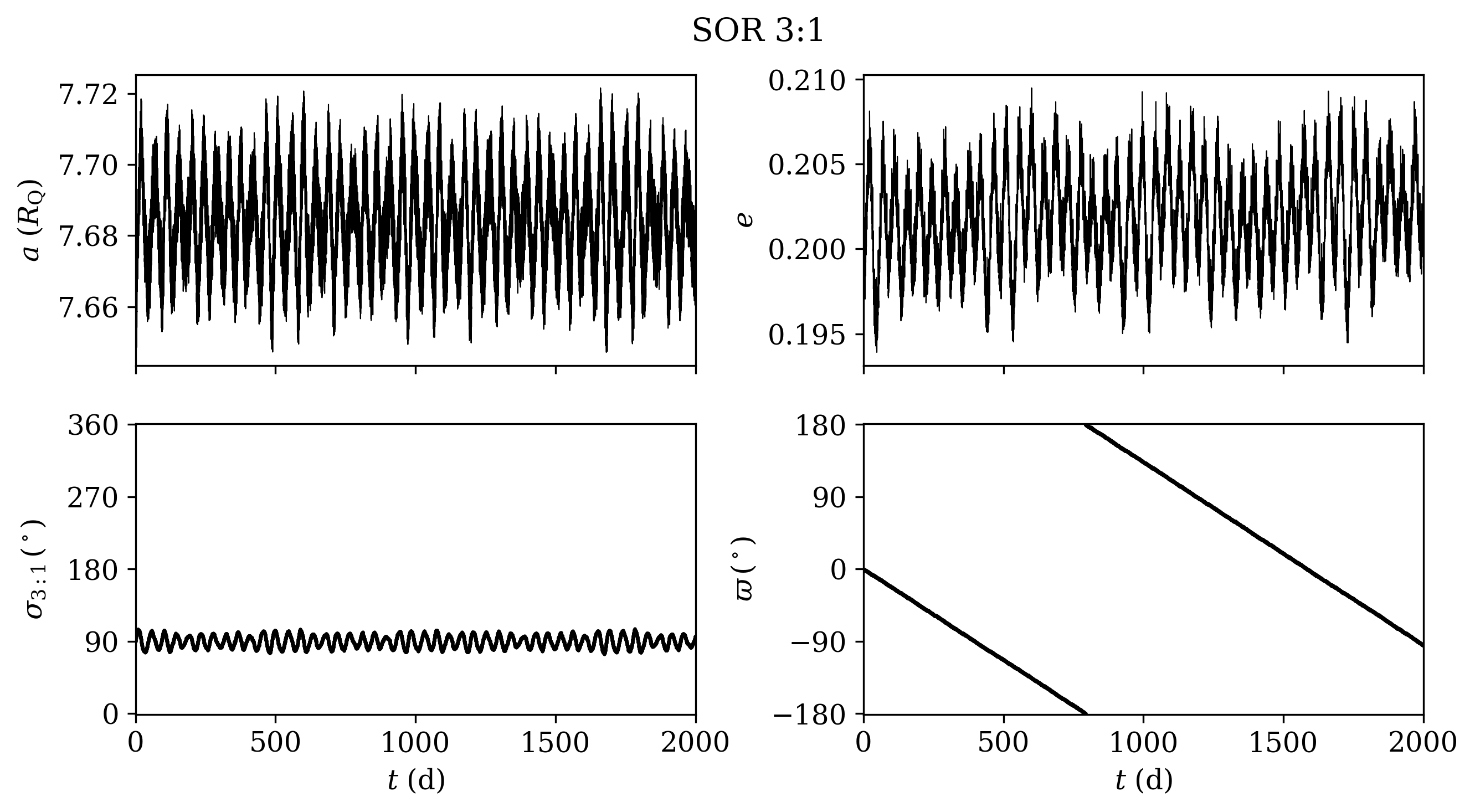}
\par\end{centering}
\begin{centering}
\includegraphics[width=0.49\textwidth]{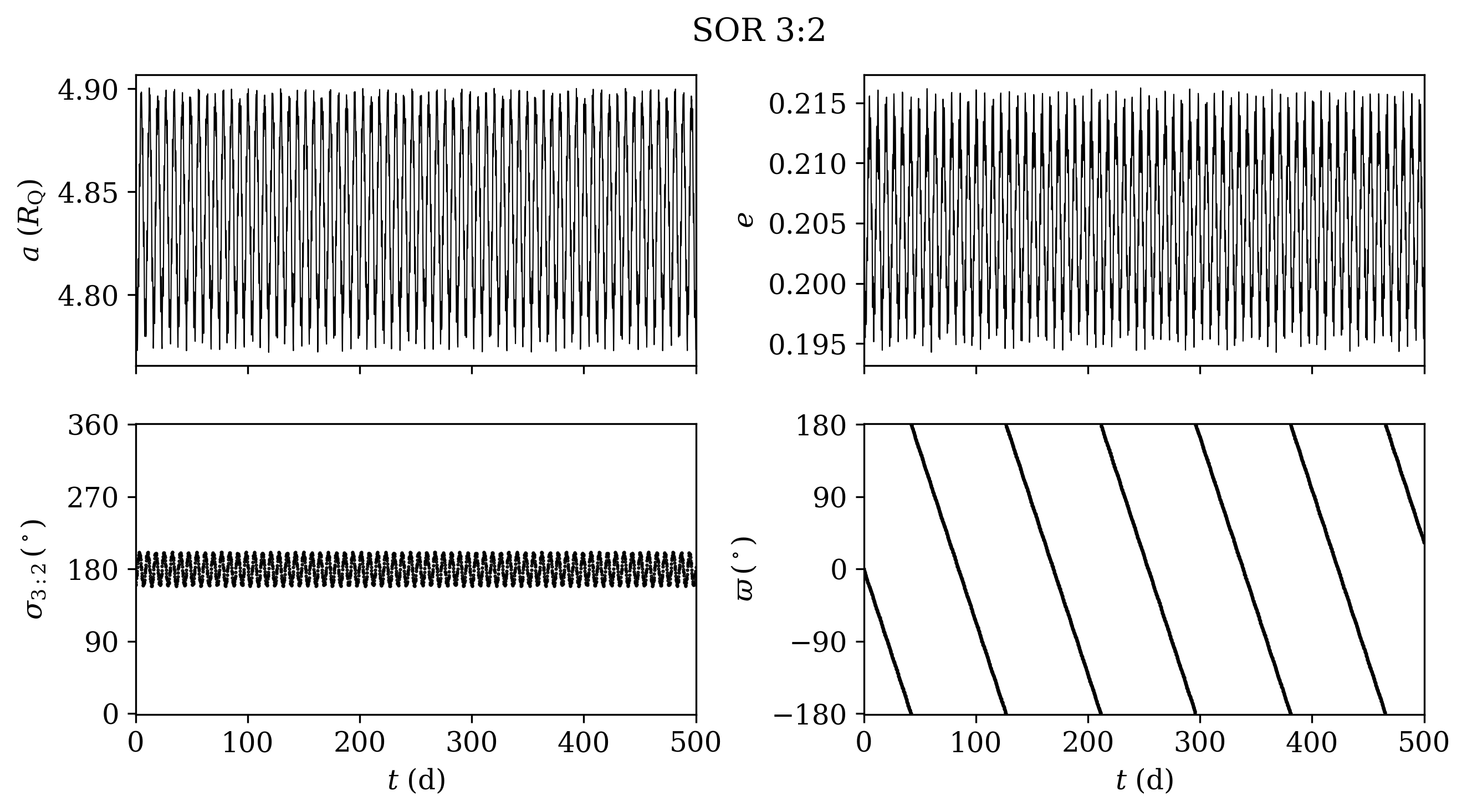}
\par\end{centering}
\centering{}\includegraphics[width=0.49\textwidth]{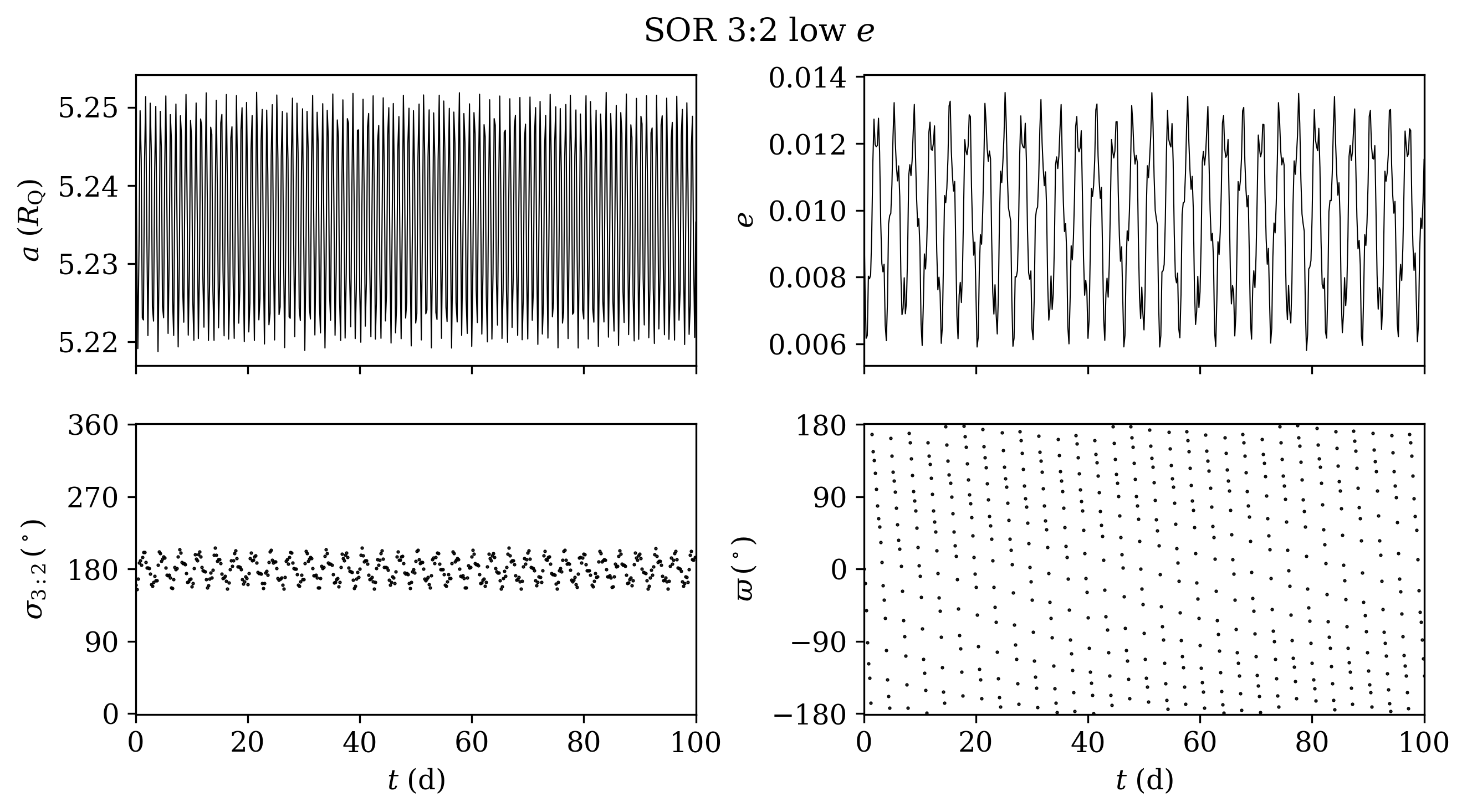}
\end{figure}

Figure \ref{fig:Examples-of-the} shows the resonant orbits at the
3:1 and 3:2 SORs. The high eccentricity orbits start with an initial
$e=0.2$. As expected, the resonant angle librates around $90^{\circ}$
and $180^{\circ}$, respectively. The libration periods, in both cases,
are in excellent agreement with the values predicted by the semianalytical
model (Fig. \ref{fig:periods}). The longitude of pericenter is always
retrograding, which appears to be a characteristic of the orbits in
the SORs. The amplitude modulation in $a,e$ that is observed in the
3:1 SOR is induced by the two satellites perturbations. 

For the particle at low eccentricity in the 3:2 SOR, the oscillation
of the resonant angle $\sigma_{3:2}$ is not a libration in the topological
sense, but rather a circulation with a forced eccentricity, usually
referred to as kinematic or paradoxical libration. This is a characteristic
of the law of structure of the resonance. In this case, the oscillation
period is much smaller that the prediction of the semianalytical model,
which is expected since the model does not reproduce the law of structure.

\begin{figure}
\caption{Examples of the orbital evolution ($a,e,\varpi,\sigma$) of two ring
particles. The particle in the outer ring is initially located at
the outer edge of the ring.}\label{fig:Examples-of-the-1}

\begin{centering}
\includegraphics[width=0.49\textwidth]{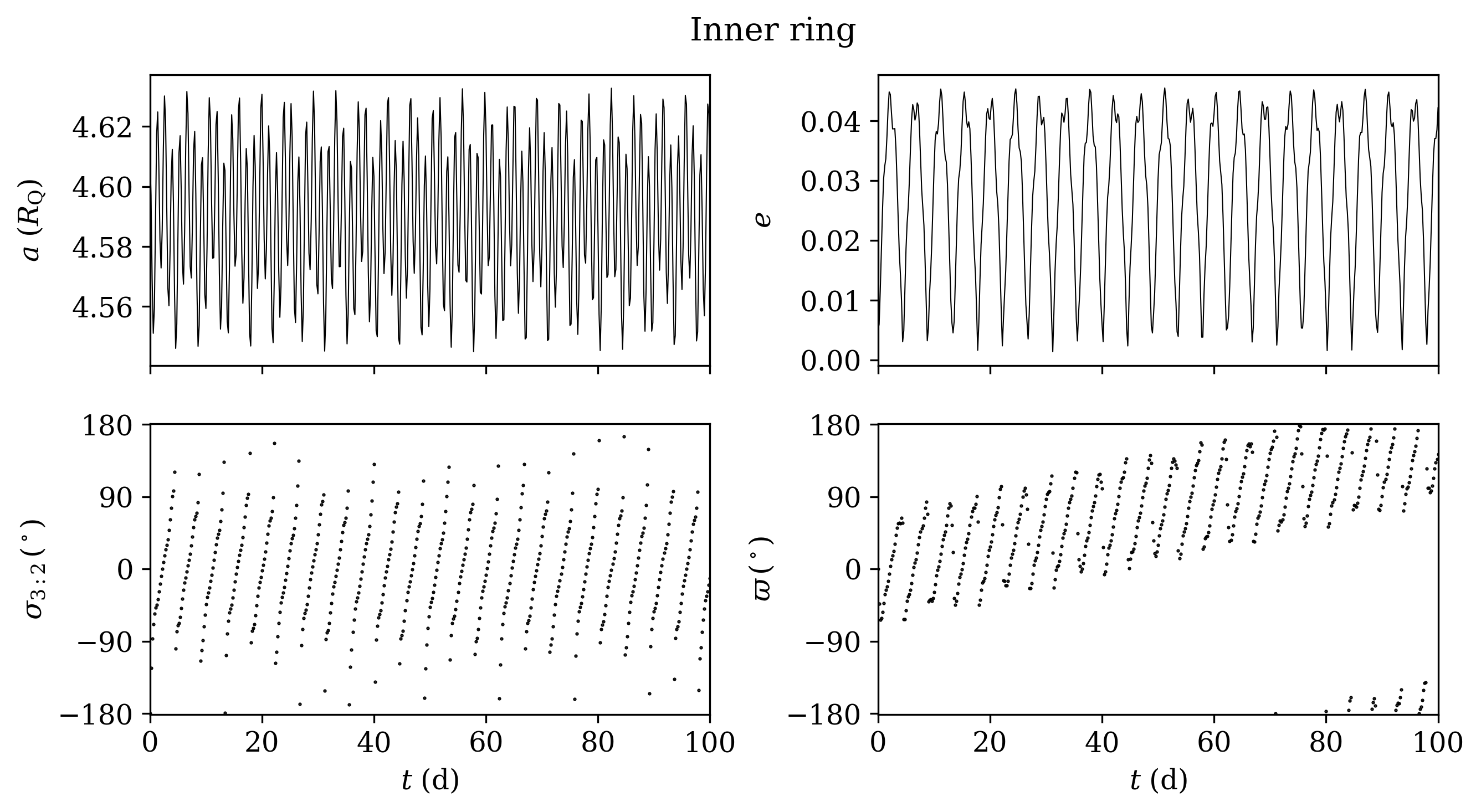}
\par\end{centering}
\centering{}\includegraphics[width=0.49\textwidth]{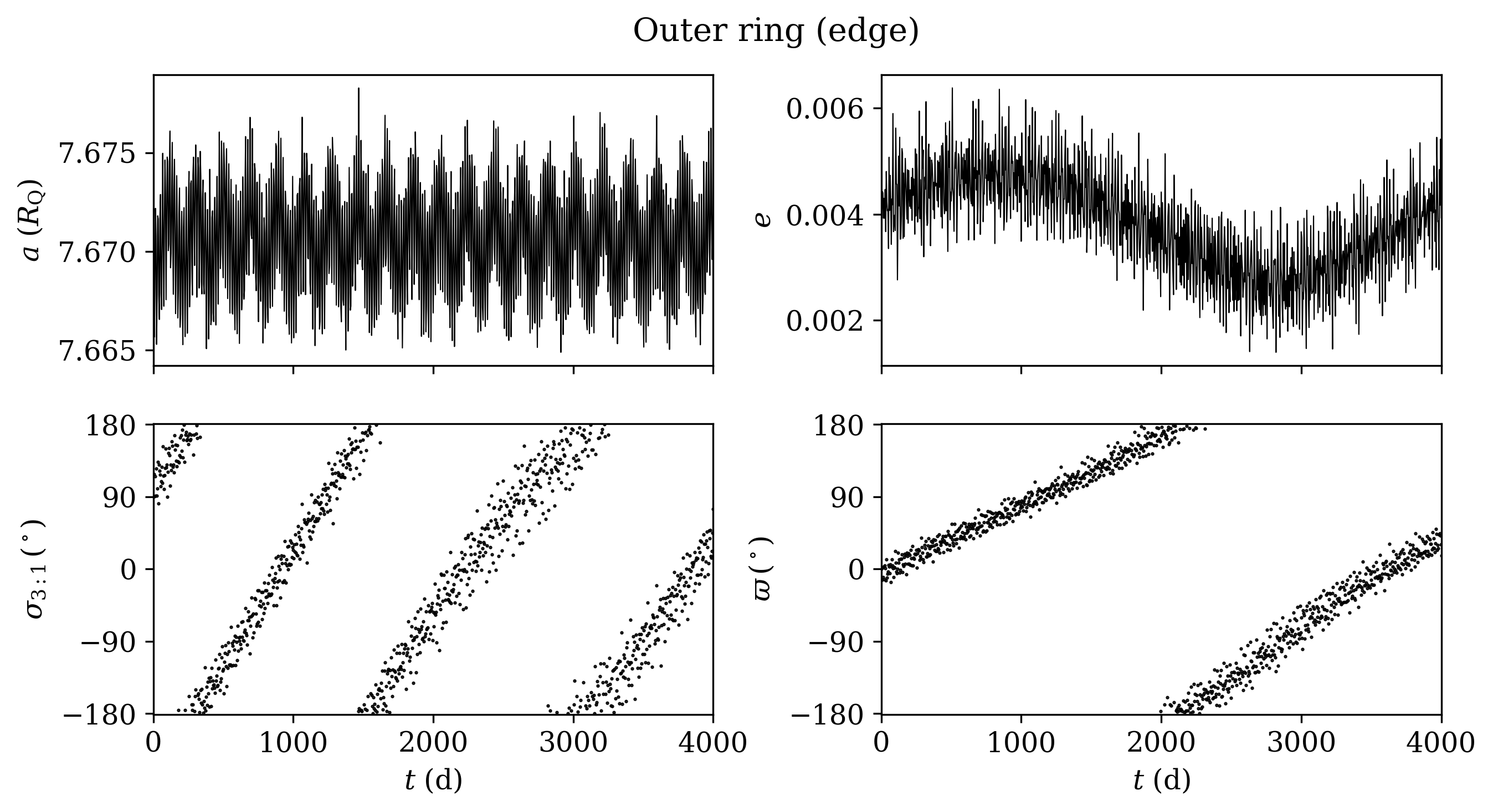}
\end{figure}

Figure \ref{fig:Examples-of-the-1} shows the orbital evolution of
test particles in the inner and outer rings, respectively. The particle
in the inner ring is clearly outside the domain of the 3:2 SOR, as
demonstrated by the fast circulation of the resonant angle. The longitude
of pericenter precess, on average, which is characteristic of the
non SOR orbits. Actually, $\varpi$ displays a short period oscillation
around an equilibrium point that is itself circulating with a much
larger period. Such short period oscillations are likely induced by
the indeterminacy of the pericenter that happens each time the eccentricity
drops to almost zero. It is noteworthy that both the semimajor axis
and the eccentricity shows large excursions, with $e$ reaching almost
0.04, as previously mentioned. The total variation in $a$ is of the
order of 40 km, which is four times the estimated width of the ring
inferred from observations \citet{2023AA...673L...4P}. Even if we
consider the averaged envelope of the oscillation amplitude, the variation
reduces to $\sim25$ km. One possible explanation for this discrepancy
is that the observations underestimated the width of the ring or its
position. Actually, if the ring radius is 50-80 km shorter than predicted,
the oscillation amplitudes in eccentricity reduce to a half, providing
a much better match to the observed width. Another possibility is
that the ring orbits are retrograde, since in this case the region
is much more stable, and typical oscillation amplitudes in $e$ do
not exceed 0.005. If neither the position nor the retrograde motion
are plausible, then additional mechanisms would need to be invoked
to keep the ring tightly confined.

For the particle in the outer ring, the situation is quite different.
This particle is located at the outer edge of the ring, with a mean
semimajor axis equal to the 3:1 SOR value ($a=3.761\,R_{Q}$). The
analysis of the resonant angle $\sigma_{3:1}$ shows that the particle
is clearly outside the resonance, although, the resonant angle actually
displays a short period oscillation around an equilibrium point that
is itself circulating with a much larger period. None of these periods
match the predictions of the semianalytical model (Fig. \ref{fig:periods}),
which are overestimated by two orders of magnitude. The longitude
of pericenter also displays a short period oscillation around an equilibrium
point that is itself precessing. The modulation in the amplitudes
of both the semimajor axis and the eccentricity is, once again, induced
by the gravitational perturbation of the two satellites. Notwithstanding,
$e$ remains very small and the maximum variation in $a$ represents
$\sim5$ km. It is worth noting that the width of the outer ring is
not well determined from the observations, and may range from 5 to
300 km \citet{2023AA...673L...4P}. While the dynamical evolution
would be enough to explain the lower estimate, a wider ring seems
to require a structure of many concentric circular orbits. This poses
constraints to the possible effects that may arise from the mutual
collisions/encounters among the ring particles.

\subsection{Mass anomaly system}\label{subsec:Mass-anomaly-system}

In this section, we analyze the case of a mass anomaly system represented
by an unequal mass binary. To setup the system, the idea is to maximize
the discrepancies with the equal mass binary potential, to better
reveal the differences between the models. We are going to consider
a binary with a mass ratio $\eta=0.267$, since this value will produce
the maximum possible discrepancy at the octupole term, according to
Eq. (\ref{eq:octupol_manom}). We will also adopt the same binary
separation as in the equal mass case ($a_{2}=0.387\,R_{Q}$), thus,
according to Eq. (\ref{eq:quadrupol_manon}), this setup will produce
a significant discrepancy at the quadrupole term too. Once
again, we recall that this is not intended to simulate the actual
dynamics around Quaoar, but rather to provide some insights on how
the dynamics differs between the equal and unequal mass dumbbells.

\begin{figure}
\caption{The widths of the main SORs in the mass anomaly system, computed with
the semianalytical model. As before, the color represents the order
of the resonance: red $q=1$, green $q=2$, blue $q=3$, gray $q=4$,
cyan $q=5.$ }\label{fig:widths-manon}

\centering{}\includegraphics[width=0.5\textwidth]{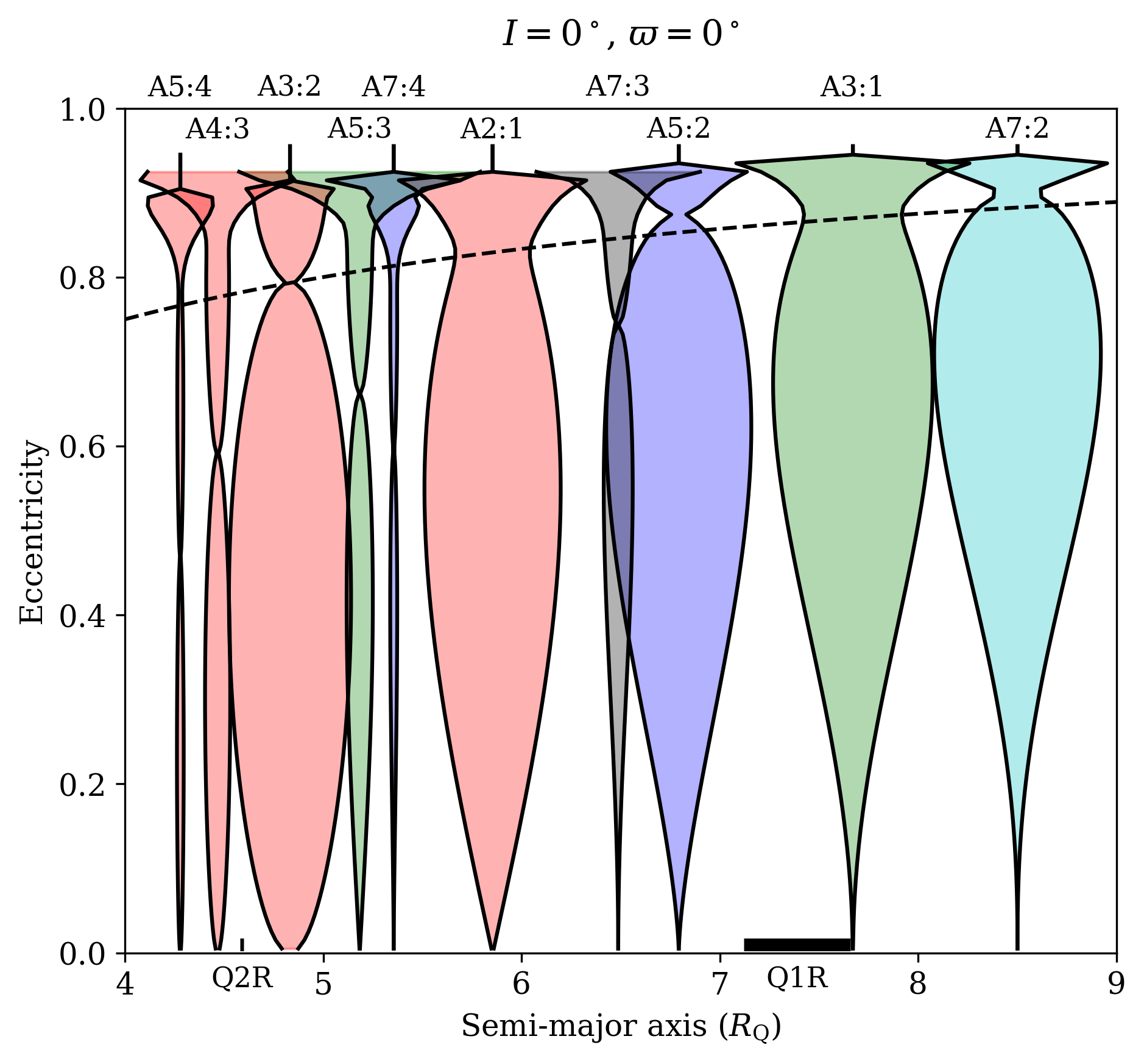}
\end{figure}

Figure \ref{fig:widths-manon} shows the SORs generated by the mass
anomaly system, calculated with the semianalytical model. This shall
be compared to Fig. \ref{fig:widths-i0}(a). Several differences are
notorious: (i) the main resonances (7:2, 3:1, 5:2, 2:1, 3:2) are about
80\% narrower than in the equal mass system, as expected from Eq.
(\ref{eq:width-scale}), and they do not longer overlap; (ii) the
resonances of degree $l=3$ become relevant, while they were extremely
weak in the equal mass system (actually, the 5:3 and 7:3 SORs were
so weak that they are not even shown in Fig. \ref{fig:widths-i0});
and (iii) resonances of degree $l=4$ do not display any appreciable
change with respect to the equal mass case. 

The explanation for this behavior relies, once again, in the harmonics
structure of the disturbing function, as discussed in Sect. \ref{subsec:The-J2-and}.
In this case, Eqs. (\ref{eq:harmonic}) and (\ref{eq:coeff}) still
hold, but now $i=2,3,4,5,\ldots$ due to the rotational asymmetry
of the potential. Therefore, for resonances with $l\geq2$, the dominant
harmonic will always have $s=1$, and it will be driven by the $l$-th
order multipole. In this case, resonances like 4:3, 5:3, 7:3 are driven
by the octupole, while in the equal mass case they were driven by
the tetrahexacontapole. On the other hand, resonances like 5:4, 7:4
are driven by the hexadecapole in both the equal and unequal mass
systems.

\begin{figure}
\caption{Location of the stable equilibrium points for the 3:1 SOR in the mass
anomaly system. The vertical dashed line indicates the collision curve.
}\label{fig:libration=000020centers}

\centering{}\includegraphics[width=0.5\textwidth]{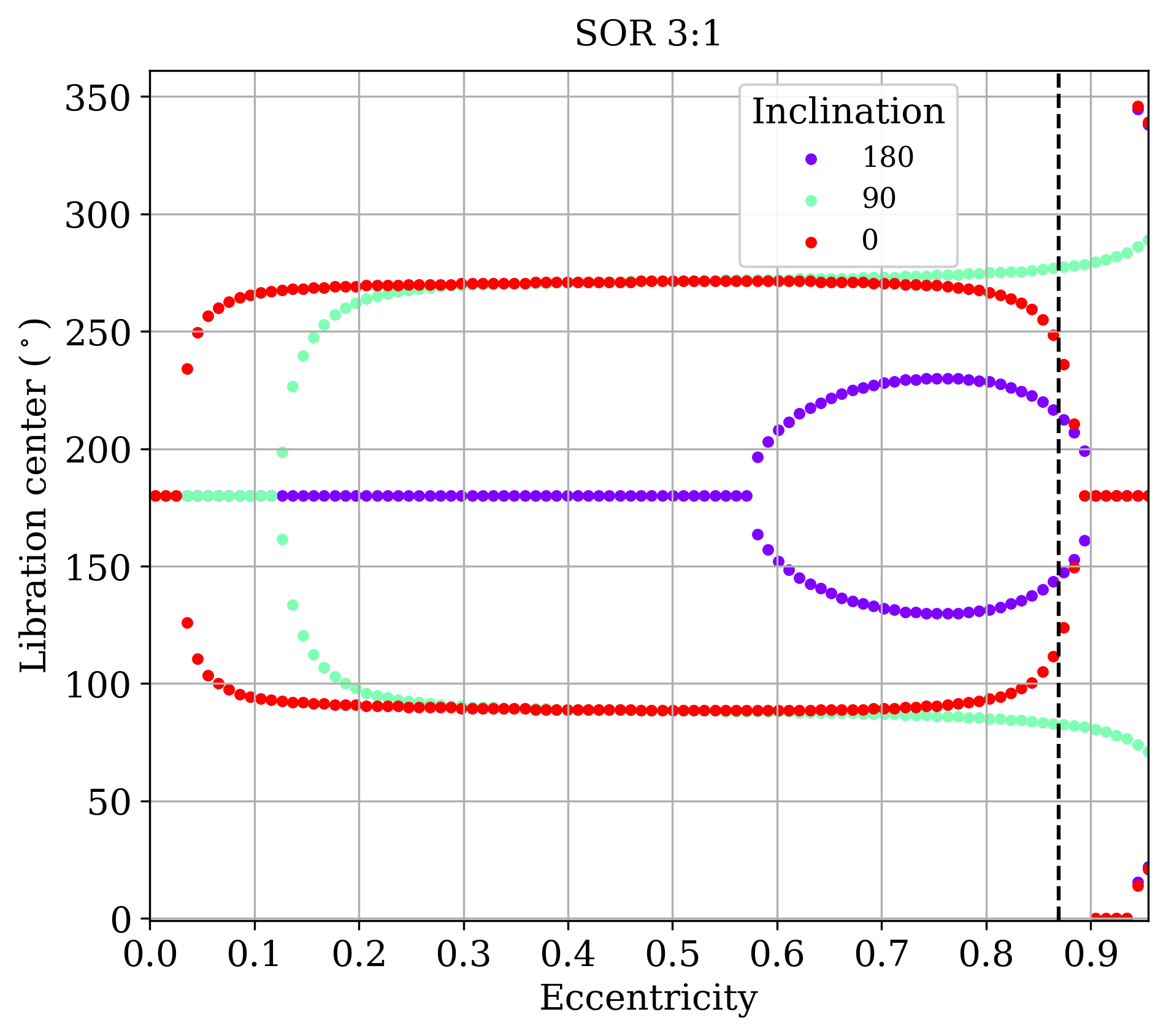}
\end{figure}

This also has implications for the location and number of equilibrium
points. Figure \ref{fig:libration=000020centers} shows the location
of the libration centers for the 3:1 SOR as a function of eccentricity
and inclination. At variance with the equal mass case, here we observe
that the stable equilibrium may bifurcate, leading to the appearance
of asymmetric librations. Again, this result is consistent with the
results of Gianuzzi et al. (2026), when their two boulders model is
configured in conjunction. This more complex structure is due to the
fact that, for resonances of degree $l=1,$ the lowest order harmonics
in the disturbing function are $\mathcal{A}_{1}^{k,1}(e)\cos\sigma+\mathcal{A}_{2}^{k,1}(e)\cos2\sigma$,
the first one driven by the octupole, and the second one driven by
the quadrupole. Therefore, the location of the stable equilibrium
results from the competition between these harmonics. For resonances
with $l\geq2$, the dominant harmonic is always $\mathcal{A}_{1}^{k,l}(e)\cos\sigma$,
thus there will be only one stable equilibrium either at $\sigma_{0}=180^{\circ}$
(for low eccentricities) or $\sigma_{0}=0^{\circ}$ (for high eccentricities).
Note that, at variance with the equal mass case, in the mass anomaly
model the 2:1 SOR is actually a first-order resonance, and the order
of any $k$:$l$ resonance will always be defined by $k-l$.

It is worth noting that all the SORs displayed in Fig. \ref{fig:widths-manon}
show a strangulation at certain eccentricities. These strangulations
are also observed in Fig. \ref{fig:widths-i0}, but they are less
conspicuous since the higher degree SORs are extremely narrow. For
the main resonances, such strangulation coincides approximately with
the collision curve, but this is not the case for the higher degree
SORs. In fact, the shrinking of the resonance width in such cases
is related to the shift from $180^{\circ}$ to $0^{\circ}$ of the
libration centers. This kind of phenomenon is also observed in MMRs
when there is a change in the topology (\citealp{2023CeMDA.135....3G};
Pan et al., 2026, in preparation).

\begin{figure}
\caption{Topology of the 3:1 SOR in the mass anomaly system, for prograde and
retrograde orbits, computed with the semianalytical model. The asymmetric
shape of the librations is evident. }\label{fig:topology-manon}

\centering{}\includegraphics[width=0.5\textwidth]{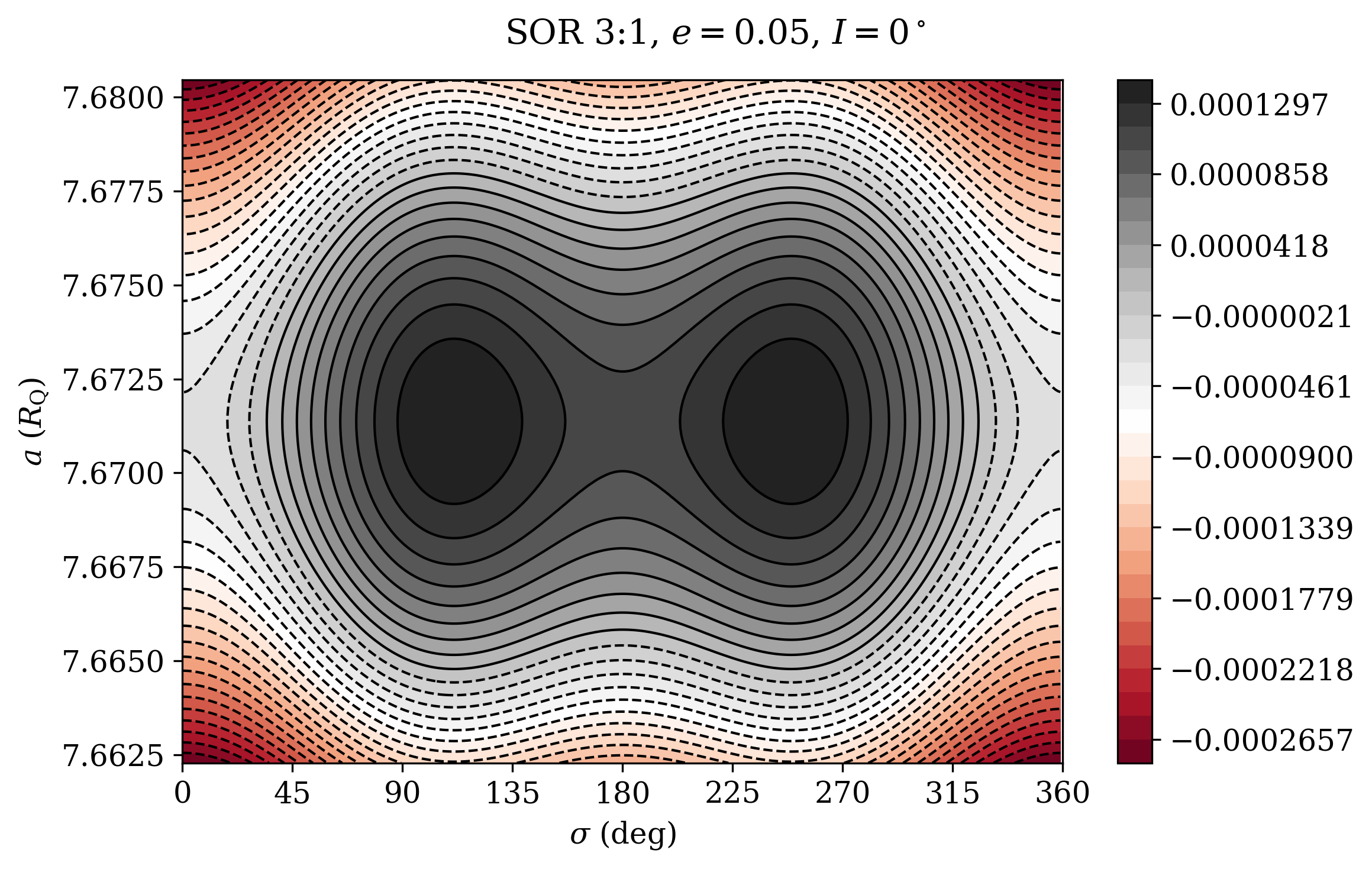}\includegraphics[width=0.5\textwidth]{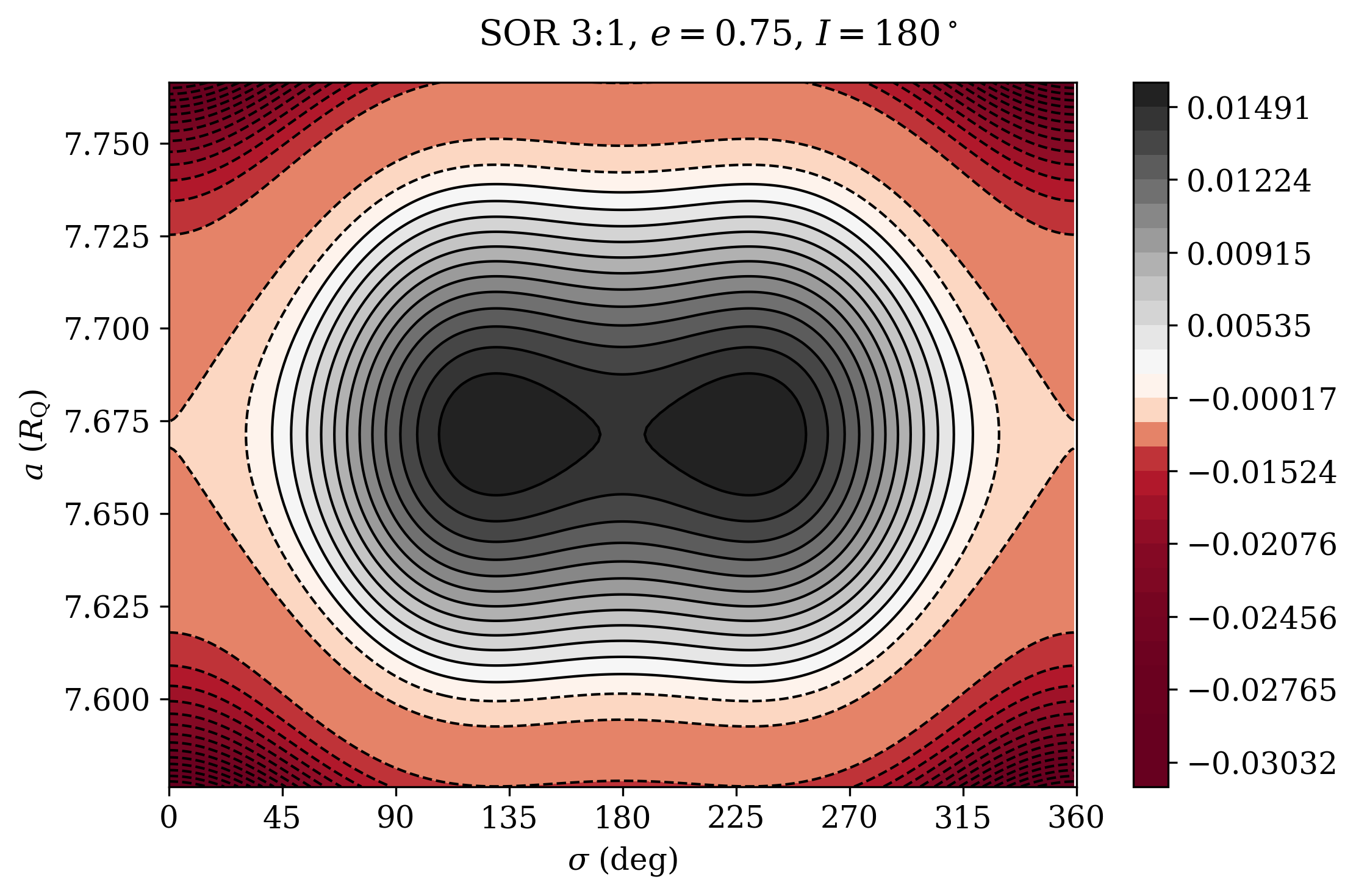}
\end{figure}

Figure \ref{fig:topology-manon} shows the topology of the 3:1 SOR
for a prograde and a retrograde orbit, at eccentricities where asymmetric
librations are present. The occurrence of asymmetric librations in
the $k$:1 SORs in the mass anomaly problem is analogous to the well
know occurrence of asymmetric librations in $p$:1 MMRs, when the
inner perturber is in a quasi circular orbit \citep{1994CeMDA..60..225B}.

For the planar case, the above results are similar
to those presented by \citet{2026A&A...708A.304B} (in particular,
their figures 3 and 4), but with some differences. While the bifurcation
here appears at $e\simeq0.03$, in their case it appears at $e\simeq0.07$.
This is partly explained because their model assumed a rotation period
of 8.8 h, while here the period is twice that value. Actually, \citet{2026A&A...708A.304B}
shows that the larger the rotation period, the smaller the bifurcation
eccentricity. Their model is also capable of detecting the kinematic
librations in the 3:1 SOR that occur at very low eccentricities ($e<0.015$),
which in our case is not possible due to the limitations of the semi-analytical
model.

\subsubsection{Dynamical maps and ring particles evolution}

\begin{figure}
\caption{Dynamical maps of the space around the Quaoar system, considering
a mass anomaly model. For a description refer to Fig. \ref{fig:dynamical-maps}.
}\label{fig:dynmaps-manon}

\centering{}\includegraphics[width=0.5\textwidth]{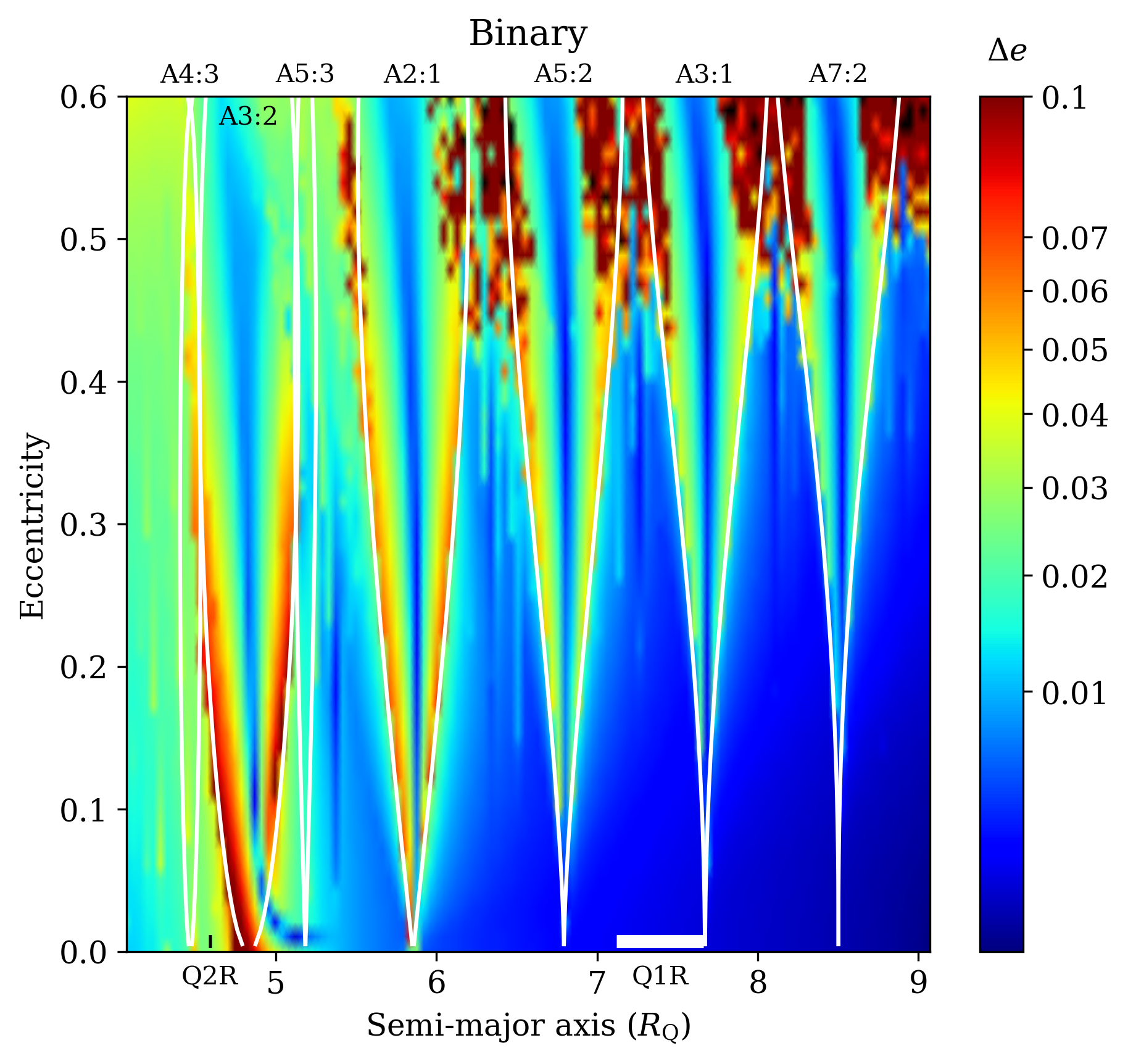}\includegraphics[width=0.5\textwidth]{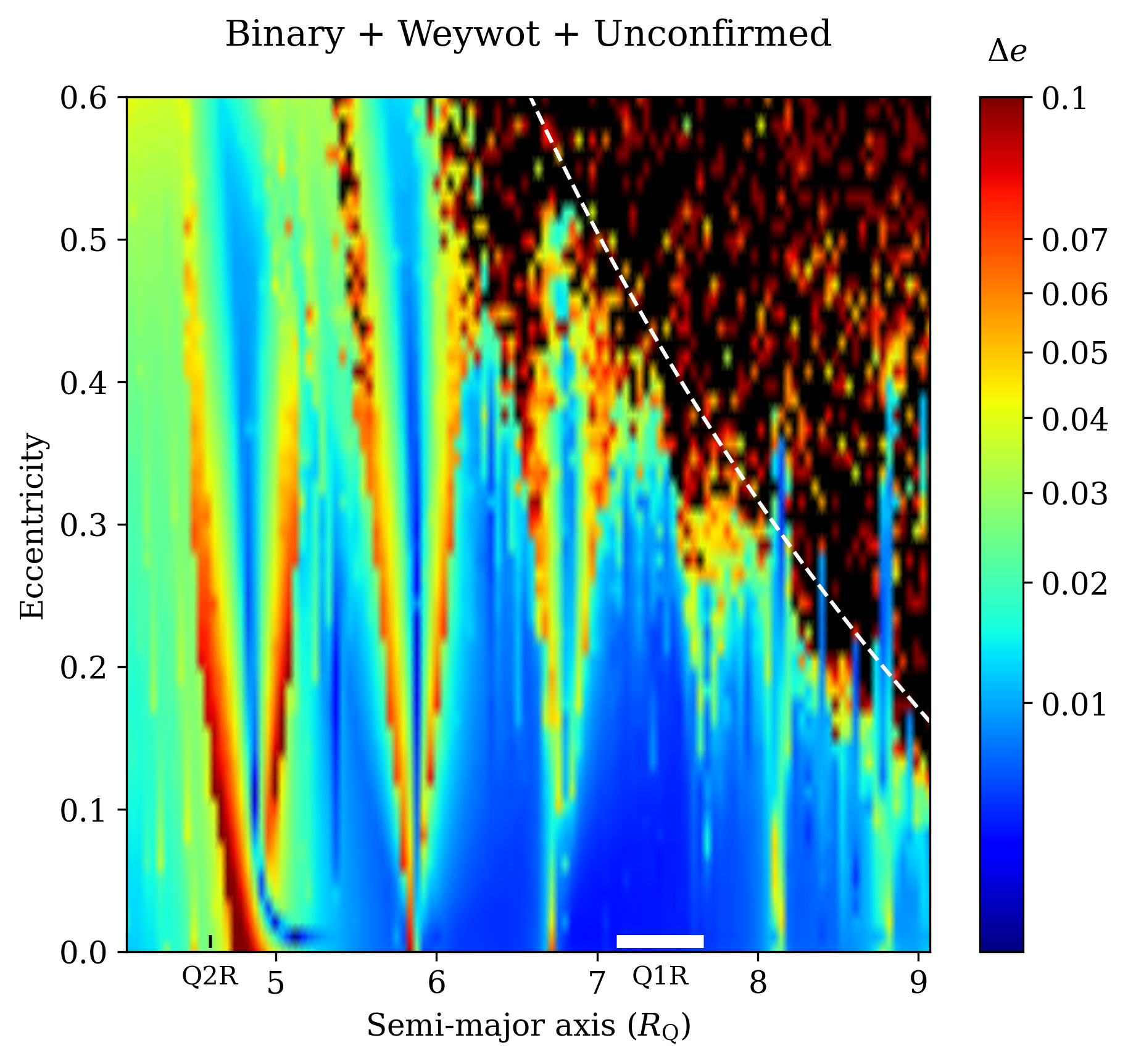}
\end{figure}

Using the same approach of the equal mass case, we constructed dynamical
maps for the mass anomaly system considering the same grid of initial
conditions. The results are presented in Fig. \ref{fig:dynmaps-manon},
which shall be compared to those of Fig. \ref{fig:dynamical-maps}.
In general, we can see that the region as a whole appears to be more
stable in this case, especially at high eccentricities. This is related
to the fact that the main SORs ($l=1,2$) are less overlapped. This
happens in spite of the appearance of SORs of degree 3, which apparently
have no significant impact in the dynamics. At low eccentricities,
no appreciable changes are observed. The outer ring is still located
at the apparently most stable region of the Quaoar system. 

It is worth noting that, in these maps, the 2:1 SOR still does not
display a law of structure at low eccentricities, which in principle
is unexpected since it is a first order resonance. However, we must
bear in mind that this map shows the case of co-planar orbits, and
similarly to the behavior observed in Fig. \ref{fig:libration=000020centers},
the bifurcation in this case occurs already at very low eccentricities
($e\lesssim0.005$). Therefore, the dominant harmonic is of order
$s=2$ and the 2:1 SOR effectively behaves as a second order resonance.
A similar behavior was observed by \citet{2026A&A...708A.304B},
who although were able to compute the loci of the law of structure
for the 2:1 SOR before the bifurcation, they did not detect any relevant
instability along it in their dynamical maps. Gianuzzi et al. (2026),
on the other hand, showed that for certain configurations of their
two boulders model, where the bifurcation occurs at larger eccentricities,
the 2:1 SOR does display a law of structure at low eccentricities,
behaving as a first-order resonance.

\begin{figure}
\caption{Examples of the orbital evolution ($a,e,\varpi,\sigma$) of two ring
particles in the mass anomaly model. The particle in the outer ring
is initially located at the outer edge of the ring.}\label{fig:Examples-manon}

\begin{centering}
\includegraphics[width=0.49\textwidth]{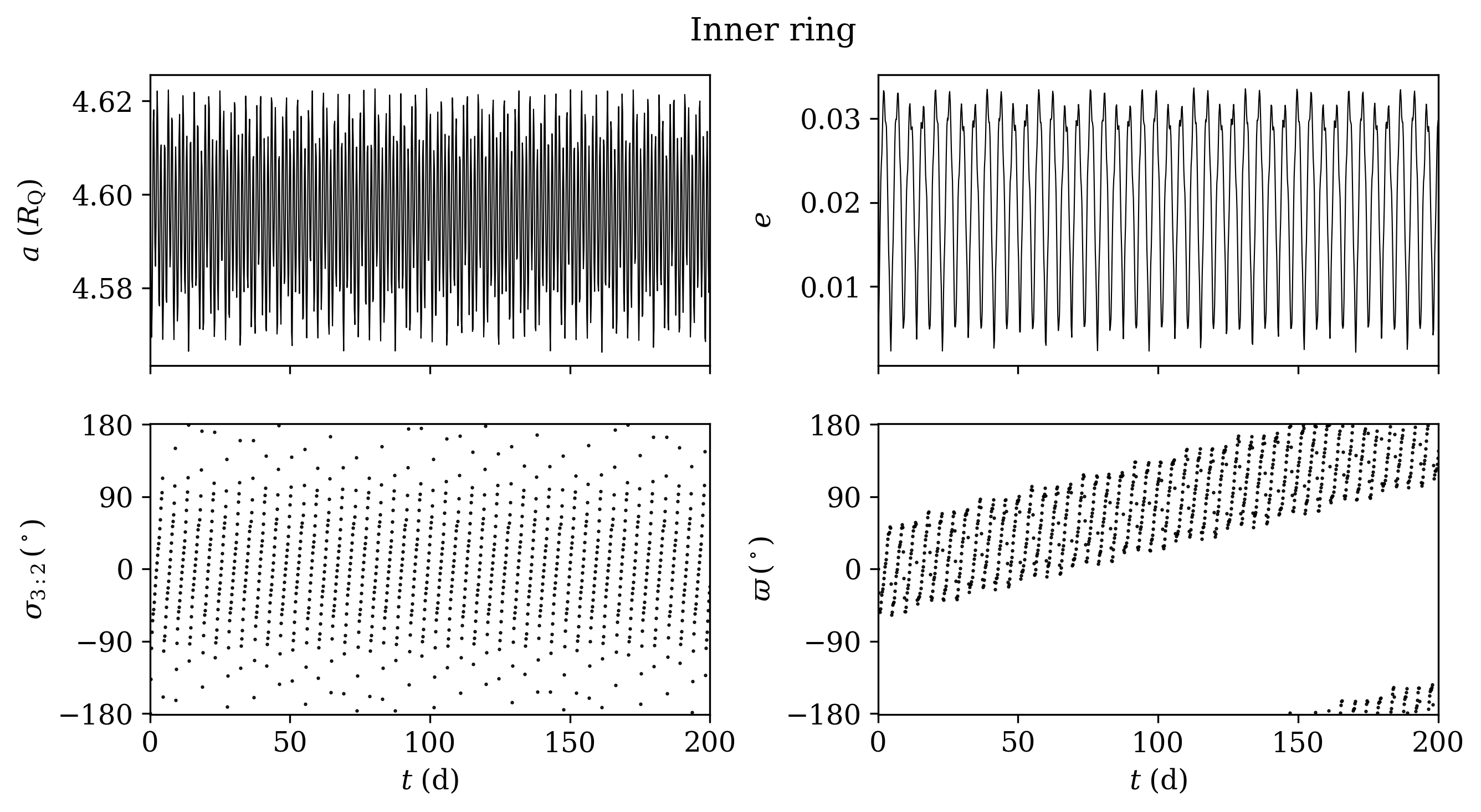}
\par\end{centering}
\centering{}\includegraphics[width=0.49\textwidth]{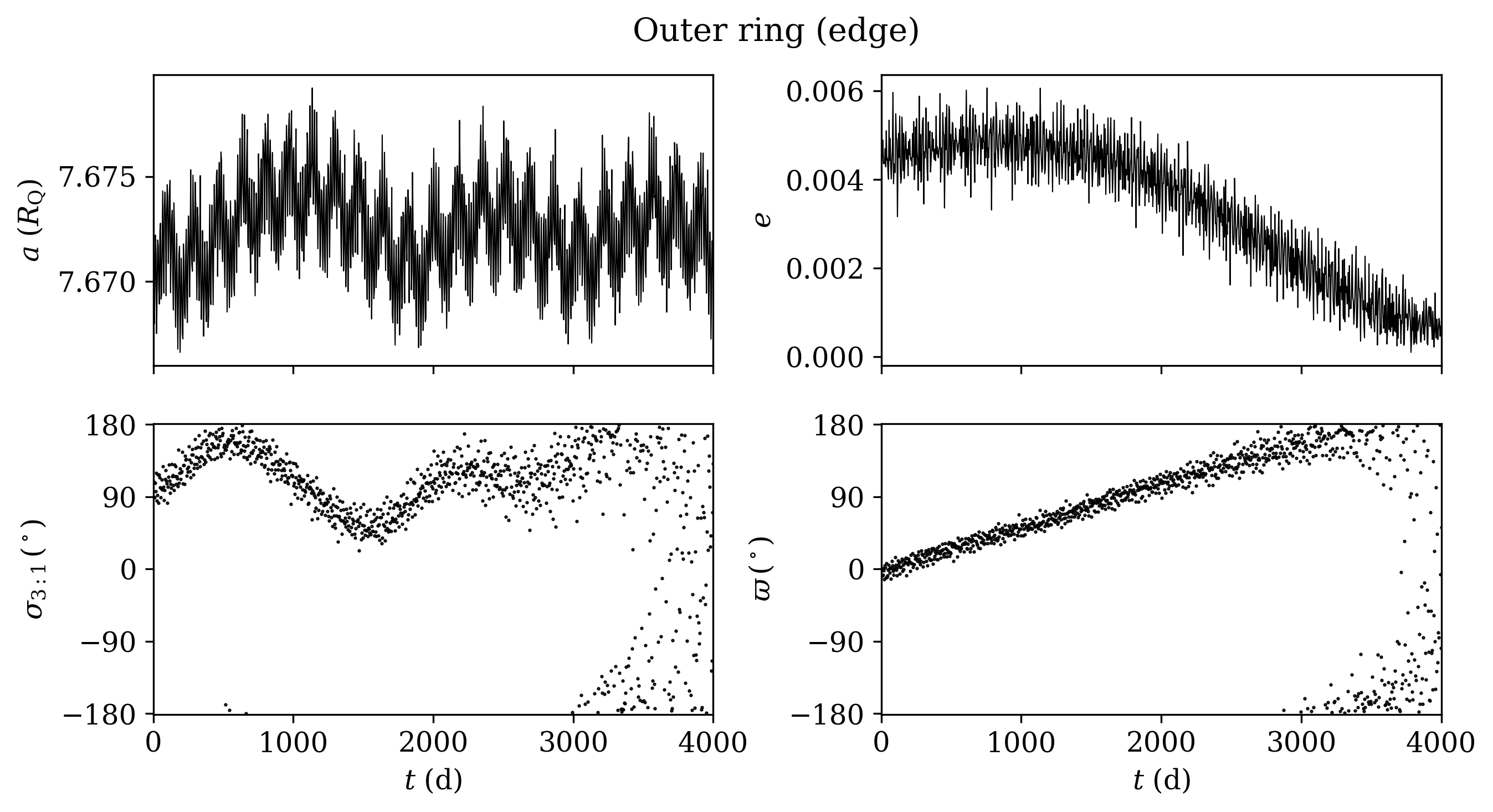}
\end{figure}

We have also analyzed the dynamical evolution of specific test particles
located in the inner and outer rings. The results are presented in
Fig. \ref{fig:Examples-manon}, and shall be compared to those of
Fig. \ref{fig:Examples-of-the-1}. For the particle in the inner ring,
the behavior is very similar as in the equal mass binary model, except
for the fact that $e$ is more limited and the oscillation in $a$
is restricted to a maximum of $\sim20$ km (but still above the estimated
width of the ring). For the outer ring particle, the evolution is
a bit different than in the equal mass binary model. This time, the
particle appears to be temporarily captures in a resonant libration,
although later it abandons this state. Once again, the modulation
in the amplitudes of $a$ and $e$ is driven by the satellites perturbation.
Nevertheless, the eccentricity remains very small, and the maximum
oscillation in $a$ still corresponds to $\sim5$ km. The proximity
to the 3:1 SOR appears to have no influence at all in the dynamics. 

\section{Conclusions}\label{sec:Conclusions}

In this work, we have presented a circumbinary approach to model the
spin-orbit resonances (SOR) that may arise when a small body orbits
a rotating irregular shaped body. The model allows to study the dynamical
behavior of these resonances in a simple way, by replacing the complex
representation of extended bodies in spherical harmonics by a direct
$N$-body representation. It also allows to replace the study of a
spin-orbit resonance by the study of a surrogate mean motion resonance.
The model was applied to the rings system around the TNO 5000 Quaoar.
Our conclusions, concerning the dynamics and structure of the SORs,
can be summarized as follows:
\begin{itemize}
\item We can state that the structure of a $k$:$l$ SORs is governed by
its degree $l$ rather than by its order $k-l$, but we must bear
in mind that for some resonance the true order is not $k-l$, but
$2(k-l)$. So, in the end, their structure is equivalent to that of
the exterior MMRs.
\item Only the low degree SORs ($l=1,2$) are relevant for the dynamics,
since these resonance are driven by the lowest order multipole terms.
Higher degree SORs ($l\geq3$) are driven by higher order multipoles
and are, in general, negligible, even when odd-order multipoles are
present.
\item In the case of a central body with rotational $\pi$-symmetry, the
SORs with $l$ odd will always display a topology with two symmetric
stable equilibrium points, while those with $l$ even will always
display only one stable equilibrium. For $l=1,2$ the locations of
these equilibria are at $\pm90^{\circ}$ and $180^{\circ}$, respectively,
and do not change with neither eccentricity, nor inclination. For
$l\geq3$ and co-planar orbits, the locations can shift depending
on eccentricity, but this does not happen for high inclination orbits.
Neither bifurcations, nor asymmetric librations occur in these cases.
\item In the case of a central body with a mass anomaly, the SORs with $l=1$
are the only ones that display a topology where bifurcation of a unique
equilibrium may lead to asymmetric librations. For SORs with $l\geq2$,
only one stable equilibrium is possible. For $l=2$, this equilibrium
is always at $180^{\circ}$in spite of the eccentricity and the inclination.
For $l\geq3$ and co-planar orbits, this equilibrium may shift from
$180^{\circ}$ to $0^{\circ}$ depending on eccentricity, but this
does not happen for high inclination orbits.
\item The width of the SORs shrinks depending on the mass ratio of the binary,
and also with the inclination of the particle. For retrograde orbits,
the SORs are extremely narrow. Retrograde orbits are expected to be
much more stable in the long term than prograde orbits. 
\item The 3:2 SOR is the only resonance to clearly display a law of structure,
either in the equal or unequal mass binary.
\item The longitude of pericenter precess for orbits that are not in SORs,
but retrogrades for orbits that are captured in SORs.
\item Systems aiming to simulate a mass anomaly need to be carefully setup,
in order to avoid reproducing a dynamics that would be generated by
an ellipsoidal body.
\item Replacing an extended body by a binary dumbbell seems to be an excellent
first order approach to the problem of the dynamics around irregular
shaped bodies.
\end{itemize}
Concerning the applications of the circumbinary model to the Quaoar
system, our conclusions can be summarized as follows:
\begin{itemize}
\item The region around the Quaoar system spanned by the two rings (between
4 and 9 Quaoar radii) is dominated by low degree SORs. The overlap
of these resonances causes some instability only at high eccentricities
($e\gtrsim0.3$).
\item The mean motion resonances with Weywot are irrelevant to the global
dynamics in that domain.
\item Mean motion resonances with the second, unconfirmed, satellite introduces
a mild instability, but the major effect of this body is the depletion
of the space at high eccentricities and large semimajor axes.
\item The outer ring is coincidentally located in the most stable region
of the whole space explored. The eccentricity of the particles in
this ring remains confined to values of $e<0.006$, and their radial
excursion is of the order of only 5 km. If the ring has an extension
of up to 300 km, as suggested by observations, it could probably consist
of a set of concentric circular orbits, with negligible radial evolution.
This puts constraints to the possible effects of the interaction among
the ring particles.
\item The inner ring is located in a region where the eccentricity of the
test particles can be excited up to values of $e\sim0.04$. The radial
excursion of the orbits can be between 20 and 40 km, which is larger
that the ring width estimated by the observations (10 km). This discrepancy
may be avoided by assuming a slightly shorter radius of the ring,
or a retrograde motion of the ring. However, if the observations are
correct and retrograde motion is not feasible, additional mechanisms
have to be invoked to keep this ring tightly confined.
\item In spite of the presence and overlap of SORs and MMRs, the space around
the Quaoar system appears to be quite stable (at least in the short
term). A central body with a mass anomaly seems to perturb less than
an equal mass binary, even when the mass anomaly system is configured
to maximize the discrepancies at quadrupole and octupole orders.
\end{itemize}
Finally, it is important to clarify that the present study is based
on a purely conservative model. In the elaboration of the above conclusions,
we are not taking into account the possible occurrence of non conservative
effects, like orbital migration by the tidal torques of the central
body, changes in the spin rate, the interaction among the ring particles,
and external effects like radiation pressure. The study has also focused
only in the short term resonant dynamics and the conclusions might
change in the long term evolution. These issues will be the subject
of future studies. 

\vspace{1cm}
I wish to thank the referee, C. Beaugé, for helpful comments and criticism.
This work has been supported by the Brazilian National Research Council
(CNPq) grant 312429/2023-1. The numerical simulations presented in
this work have been performed at the National Observatory Center for
Data Processing (CPDON).

\appendix

\section{Ellipsoid equivalence in the long term evolution}

\begin{figure}
\caption{Long term dynamical maps of the space around the
Quaoar system, considering the equal mass circumbinary model, a prolate
ellipsoid including the hexadecapole term, and a triaxial ellipsoid.
In all cases, the SOR widths are computed with the semi-analytical
circumbinary model. }\label{fig:ellipsoid-equiv}

\begin{centering}
\includegraphics[width=0.5\textwidth]{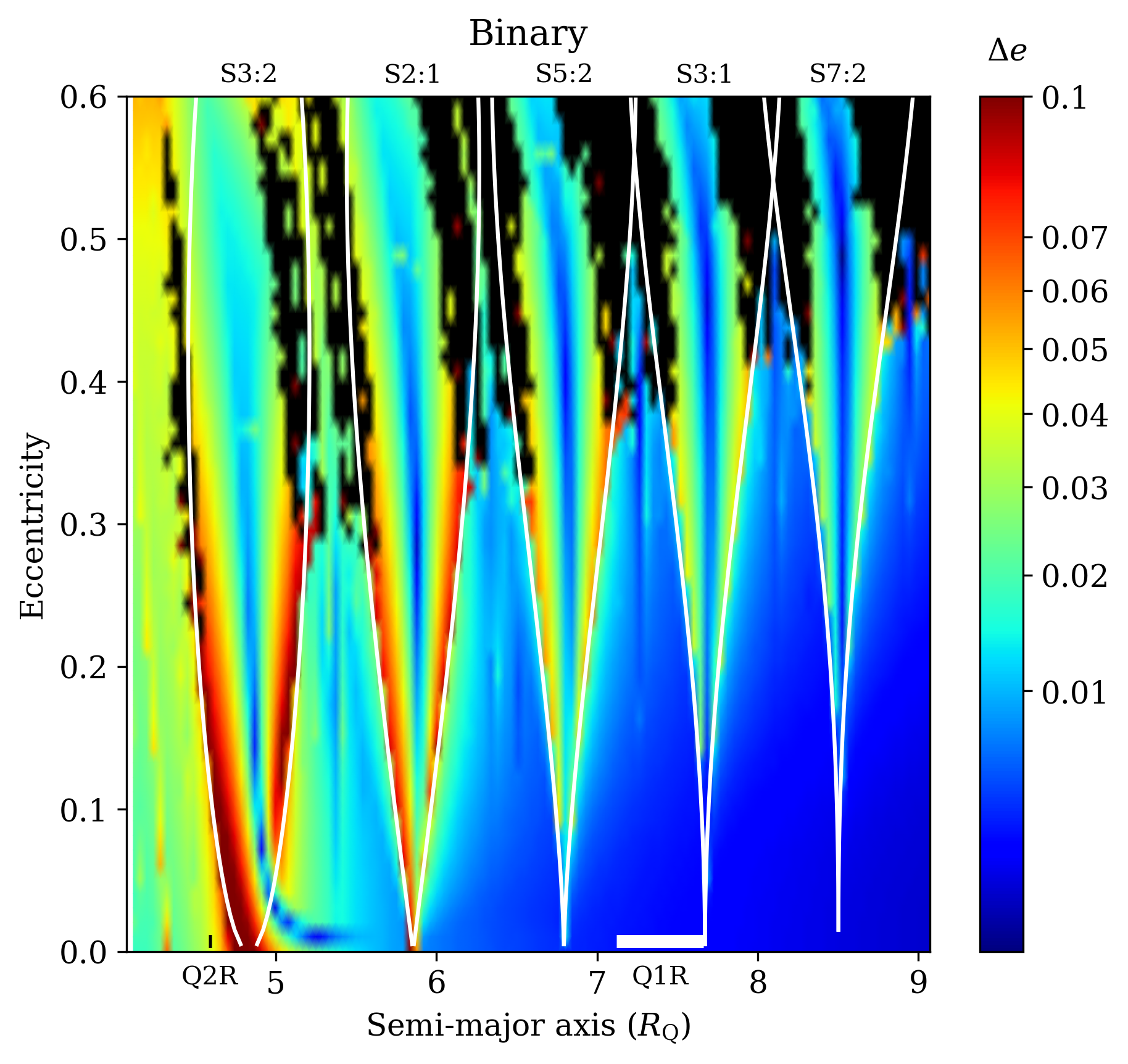}\includegraphics[width=0.5\textwidth]{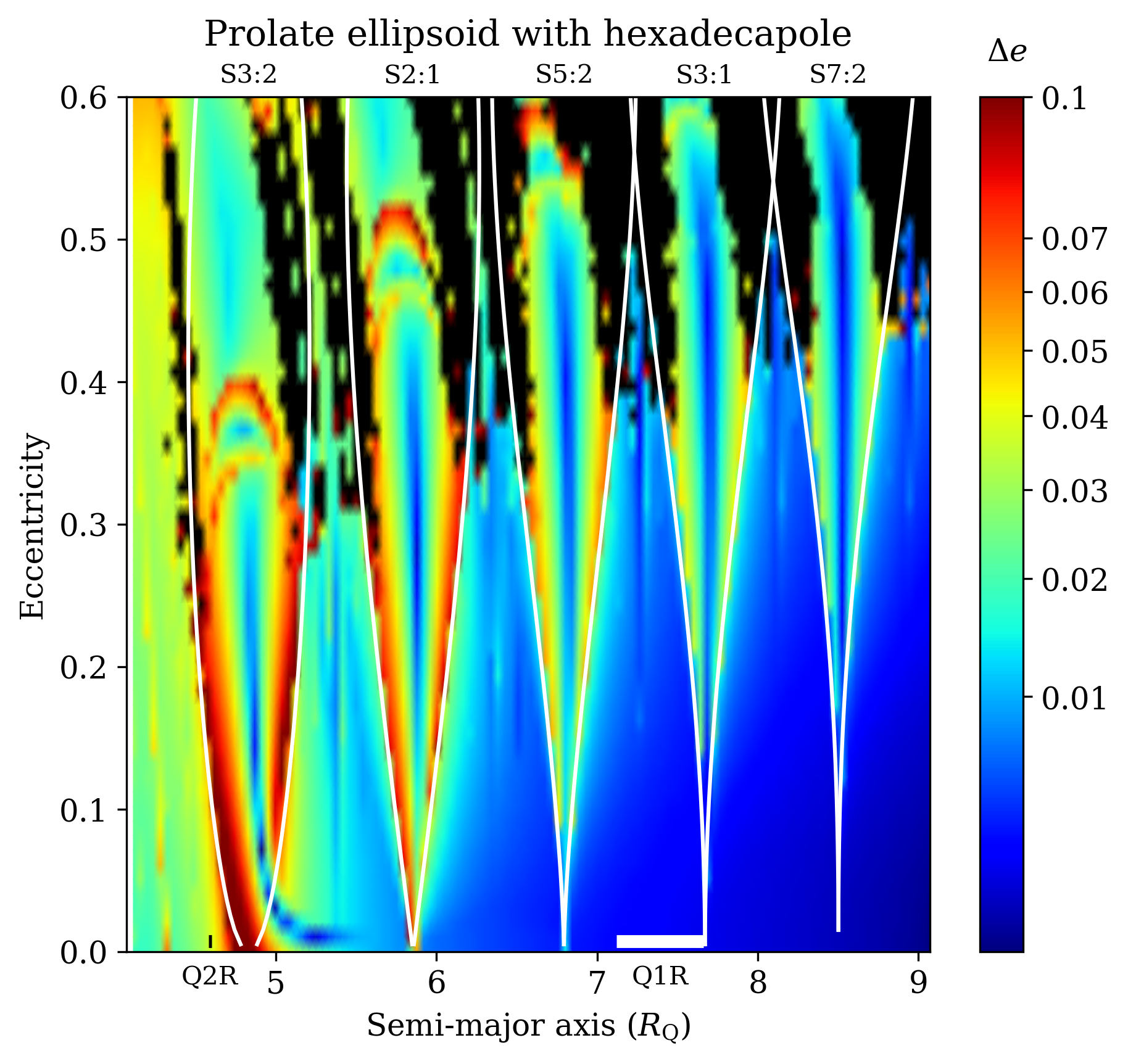}
\par\end{centering}
\centering{}\includegraphics[width=0.5\textwidth]{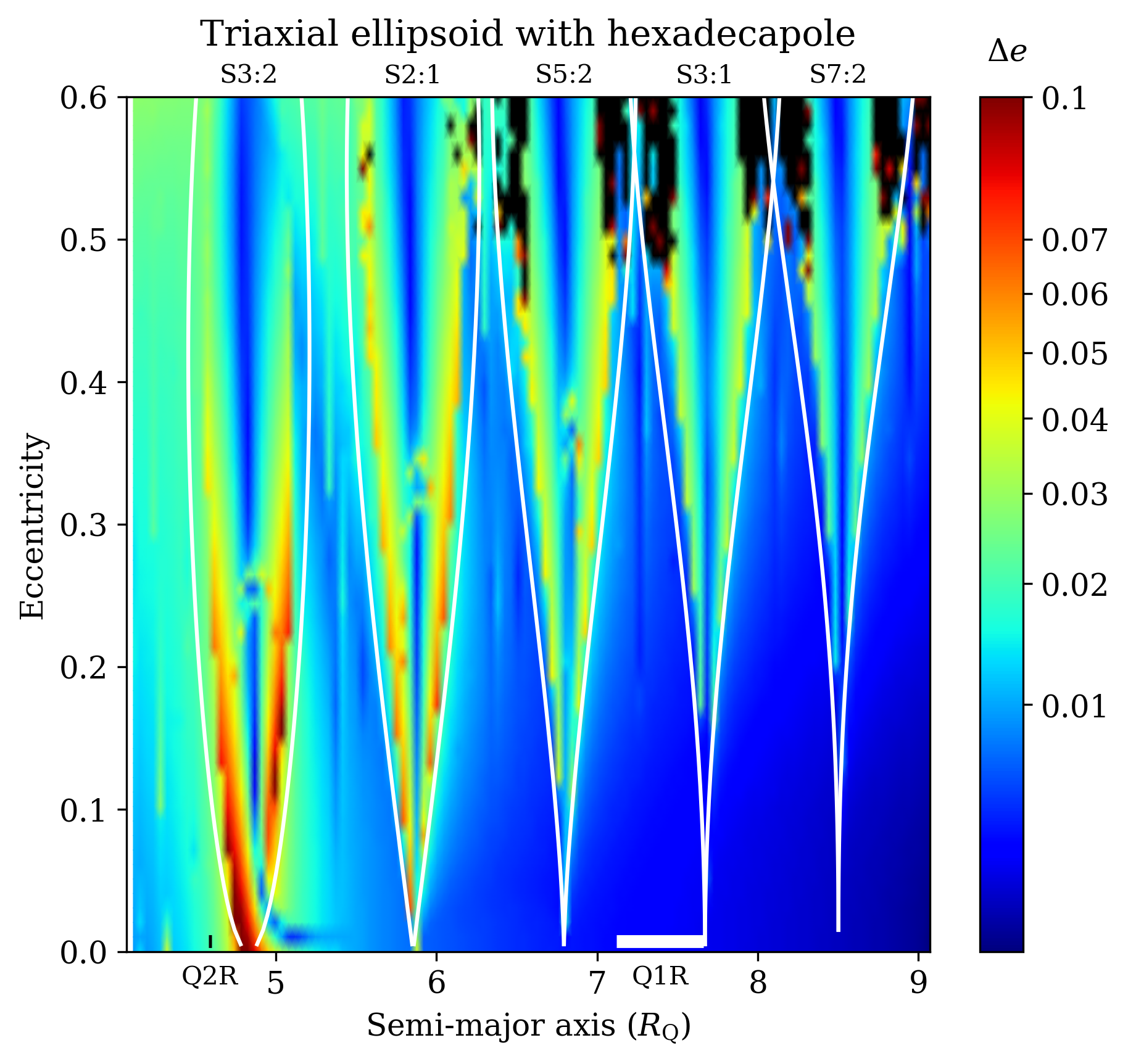}
\end{figure}

In order to address the differences in the long term
evolution of the test particles between the circumbinary approach
and the ellipsoid potential, we produced a series of dynamical maps
spanning a total time of 50,000 days, ten times larger that the time
span used in Fig. \ref{fig:dynamical-maps}. We considered the circumbinary
model, without the satellites, and a prolate and triaxial ellipsoid
potentials up to the hexadecapole term. The results are shown in Fig.
\ref{fig:ellipsoid-equiv}.

The first panel should be compared to the first panel
of Fig. \ref{fig:dynamical-maps}. We can see that there is no significant
differences between the two simulations; only at $e\gtrsim0.3$ the
depletion in the resonance overlap region is more evident. The prolate
ellipsoid model does not display any significant differences either,
only inside the resonances (especially the 3:2 and 2:1 SORs) it appears
a more chaotic structure possible associated to secular resonances
or secondary resonances, which should be driven by the hexadecapole.
Recall that the binary was tuned up to mimic the quadrupole of the
prolate ellipsoid. Thus, we may conclude that the hexadecople error
estimated by Eq. (\ref{eq:error-hexa}) is not relevant, in general,
to the long term dynamics.

The situation is completely different concerning
the triaxial ellipsoid model. Recall that it is not possible to tune
up the binary to mimic the quadrupole of a triaxial ellipsoid, which
generates an error estimated by Eq. (\ref{eq:error-tri}). Surprisingly,
a triaxial ellipsoid produces a map that is globally more stable that
the others. Not only the maximum eccentricity variations are suppressed
by a factor of $\sim2$, but also the widths of the SORs are $\sim80\%$
narrower. This means that triaxiality can be crucial for the long
term evolution of orbits around an irregular shaped body, and poses
a relevant limitation to the circumbinary approach.

\bibliographystyle{aasjournal}

\begin{thebibliography}{}
\expandafter\ifx\csname natexlab\endcsname\relax\def\natexlab#1{#1}\fi
\providecommand{\url}[1]{\href{#1}{#1}}
\providecommand{\dodoi}[1]{doi:~\href{http://doi.org/#1}{\nolinkurl{#1}}}
\providecommand{\doeprint}[1]{\href{http://ascl.net/#1}{\nolinkurl{http://ascl.net/#1}}}
\providecommand{\doarXiv}[1]{\href{https://arxiv.org/abs/#1}{\nolinkurl{https://arxiv.org/abs/#1}}}

\bibitem[{{Beaug{\'e}}(1994)}]{1994CeMDA..60..225B}
{Beaug{\'e}}, C. 1994, \cemda, 60, 225, \dodoi{10.1007/BF00693323}

\bibitem[{{Beaug{\'e}} {et~al.}(2026){Beaug{\'e}}, {Gianuzzi}, {Cerioni},
  {Zoppetti}, {Leiva}, \& {Tr{\'o}golo}}]{2026A&A...708A.304B}
{Beaug{\'e}}, C., {Gianuzzi}, E., {Cerioni}, M., {et~al.} 2026, \aap, 708,
  A304, \dodoi{10.1051/0004-6361/202558558}

\bibitem[{{Beaug{\'e}} {et~al.}(2025){Beaug{\'e}}, {Gianuzzi}, {Tr{\'o}golo},
  {Leiva}, {Zoppetti}, \& {Cerioni}}]{2025AA...702L..15B}
{Beaug{\'e}}, C., {Gianuzzi}, E., {Tr{\'o}golo}, N., {et~al.} 2025, \aap, 702,
  L15, \dodoi{10.1051/0004-6361/202556383}

\bibitem[{{Braga-Ribas} {et~al.}(2026){Braga-Ribas}, {Pereira}, {Sicardy},
  {Morgado}, {Ortiz}, {Madeira}, \& {Margoti}}]{2026ApJ...999L..39B}
{Braga-Ribas}, F., {Pereira}, C.~L., {Sicardy}, B., {et~al.} 2026, \apjl, 999,
  L39, \dodoi{10.3847/2041-8213/ae4751}

\bibitem[{{Braga-Ribas} {et~al.}(2025){Braga-Ribas}, {Vachier}, {Desmars},
  {Margoti}, \& {Sicardy}}]{2025RSPTA.38340200B}
{Braga-Ribas}, F., {Vachier}, F., {Desmars}, J., {Margoti}, G., \& {Sicardy},
  B. 2025, Phil. Trans. Royal Soc. London Ser. A, 383, 20240200,
  \dodoi{10.1098/rsta.2024.0200}

\bibitem[{{Branham}(1990)}]{1990CeMDA..49..209B}
{Branham}, Jr., R.~L. 1990, \cemda, 49, 209, \dodoi{10.1007/BF00050715}

\bibitem[{{Celletti} \& {Chierchia}(2000)}]{2000CeMDA..76..229C}
{Celletti}, A., \& {Chierchia}, L. 2000, \cemda, 76, 229,
  \dodoi{10.1023/A:1008341317257}

\bibitem[{{Celletti} {et~al.}(2007){Celletti}, {Froeschl{\'e}}, \&
  {Lega}}]{2007PSS...55..889C}
{Celletti}, A., {Froeschl{\'e}}, C., \& {Lega}, E. 2007, \planss, 55, 889,
  \dodoi{10.1016/j.pss.2006.11.023}

\bibitem[{{Chambers}(2010)}]{2010ASSL..366..239C}
{Chambers}, J.~E. 2010, in Astrophysics and Space Science Library, Vol. 366,
  Planets in Binary Star Systems, ed. N.~{Haghighipour}, 239,
  \dodoi{10.1007/978-90-481-8687-7_9}

\bibitem[{{Correia} \& {Delisle}(2019)}]{2019AA...630A.102C}
{Correia}, A. C.~M., \& {Delisle}, J.-B. 2019, \aap, 630, A102,
  \dodoi{10.1051/0004-6361/201936336}

\bibitem[{{Danby}(1962)}]{1962fcm..book.....D}
{Danby}, J.~M. 1962, {Fundamentals of celestial mechanics} (New York:
  Macmillan)

\bibitem[{{Gallardo}(2020)}]{2020CeMDA.132....9G}
{Gallardo}, T. 2020, \cemda, 132, 9, \dodoi{10.1007/s10569-019-9948-7}

\bibitem[{{Gallardo} {et~al.}(2021){Gallardo}, {Beaug{\'e}}, \&
  {Giuppone}}]{2021AA...646A.148G}
{Gallardo}, T., {Beaug{\'e}}, C., \& {Giuppone}, C.~A. 2021, \aap, 646, A148,
  \dodoi{10.1051/0004-6361/202039764}

\bibitem[{{Giuppone} {et~al.}(2023){Giuppone}, {Rodr{\'\i}guez}, {Alencastro},
  {Roig}, \& {Gallardo}}]{2023CeMDA.135....3G}
{Giuppone}, C., {Rodr{\'\i}guez}, A., {Alencastro}, V., {Roig}, F., \&
  {Gallardo}, T. 2023, \cemda, 135, 3, \dodoi{10.1007/s10569-022-10112-5}

\bibitem[{{Goldreich} \& {Peale}(1966)}]{1966AJ.....71..425G}
{Goldreich}, P., \& {Peale}, S. 1966, \aj, 71, 425, \dodoi{10.1086/109947}

\bibitem[{{Goldreich} \& {Peale}(1967)}]{1967AJ.....72..662G}
---. 1967, \aj, 72, 662, \dodoi{10.1086/110289}

\bibitem[{{Holman} \& {Wiegert}(1999)}]{1999AJ....117..621H}
{Holman}, M.~J., \& {Wiegert}, P.~A. 1999, \aj, 117, 621,
  \dodoi{10.1086/300695}

\bibitem[{{Jackson}(1999)}]{Jackson1999}
{Jackson}, J.~D. 1999, {Classical Electrodynamics (3rd. ed.)} (New York: John
  Wiley \& Sons)

\bibitem[{{Madeira} {et~al.}(2022){Madeira}, {Giuliatti Winter}, {Ribeiro}, \&
  {Winter}}]{2022MNRAS.510.1450M}
{Madeira}, G., {Giuliatti Winter}, S.~M., {Ribeiro}, T., \& {Winter}, O.~C.
  2022, \mnras, 510, 1450, \dodoi{10.1093/mnras/stab3552}

\bibitem[{{Margoti}(2024)}]{Margoti2024}
{Margoti}, G. 2024, PhD thesis, Universidade Tecnológica Federal do Paraná

\bibitem[{{Morgado} {et~al.}(2023){Morgado}, {Sicardy}, {Braga-Ribas}, {Ortiz},
  {Salo}, {Vachier}, {Desmars}, {Pereira}, {Santos-Sanz}, {Sfair}, {de
  Santana}, {Assafin}, {Vieira-Martins}, {Gomes-J{\'u}nior}, {Margoti},
  {Dhillon}, {Fern{\'a}ndez-Valenzuela}, {Broughton}, {Bradshaw}, {Langersek},
  {Benedetti-Rossi}, {Souami}, {Holler}, {Kretlow}, {Boufleur}, {Camargo},
  {Duffard}, {Beisker}, {Morales}, {Lecacheux}, {Rommel}, {Herald}, {Benz},
  {Jehin}, {Jankowsky}, {Marsh}, {Littlefair}, {Bruno}, {Pagano}, {Brandeker},
  {Collier-Cameron}, {Flor{\'e}n}, {Hara}, {Olofsson}, {Wilson}, {Benkhaldoun},
  {Busuttil}, {Burdanov}, {Ferrais}, {Gault}, {Gillon}, {Hanna}, {Kerr},
  {Kolb}, {Nosworthy}, {Sebastian}, {Snodgrass}, {Teng}, \& {de
  Wit}}]{2023Natur.614..239M}
{Morgado}, B.~E., {Sicardy}, B., {Braga-Ribas}, F., {et~al.} 2023, \nat, 614,
  239, \dodoi{10.1038/s41586-022-05629-6}

\bibitem[{{Murray} \& {Dermott}(1999)}]{1999ssd..book.....M}
{Murray}, C.~D., \& {Dermott}, S.~F. 1999, {Solar System Dynamics} (Cambridge:
  Cambridge University Press), \dodoi{10.1017/CBO9781139174817}

\bibitem[{{Pereira} {et~al.}(2023){Pereira}, {Sicardy}, {Morgado},
  {Braga-Ribas}, {Fern{\'a}ndez-Valenzuela}, {Souami}, {Holler}, {Boufleur},
  {Margoti}, {Assafin}, {Ortiz}, {Santos-Sanz}, {Epinat}, {Kervella},
  {Desmars}, {Vieira-Martins}, {Kilic}, {Gomes J{\'u}nior}, {Camargo},
  {Emilio}, {Vara-Lubiano}, {Kretlow}, {Albert}, {Alcock}, {Ball}, {Bender},
  {Buie}, {Butterfield}, {Camarca}, {Castro-Chac{\'o}n}, {Dunford}, {Fisher},
  {Gamble}, {Geary}, {Gnilka}, {Green}, {Hartman}, {Huang}, {Januszewski},
  {Johnston}, {Kagitani}, {Kamin}, {Kavelaars}, {Keller}, {de Kleer}, {Lehner},
  {Luken}, {Marchis}, {Marlin}, {McGregor}, {Nikitin}, {Nolthenius}, {Patrick},
  {Redfield}, {Rengstorf}, {Reyes-Ruiz}, {Seccull}, {Skrutskie}, {Smith},
  {Sproul}, {Stephens}, {Szentgyorgyi}, {S{\'a}nchez-Sanju{\'a}n}, {Tatsumi},
  {Verbiscer}, {Wang}, {Yoshida}, {Young}, \& {Zhang}}]{2023AA...673L...4P}
{Pereira}, C.~L., {Sicardy}, B., {Morgado}, B.~E., {et~al.} 2023, \aap, 673,
  L4, \dodoi{10.1051/0004-6361/202346365}

\bibitem[{{Proudfoot} {et~al.}(2025{\natexlab{a}}){Proudfoot}, {Grundy},
  {Ragozzine}, \& {Fern{\'a}ndez-Valenzuela}}]{2025PSJ.....6..285P}
{Proudfoot}, B., {Grundy}, W., {Ragozzine}, D., \& {Fern{\'a}ndez-Valenzuela},
  E. 2025{\natexlab{a}}, \psj, 6, 285, \dodoi{10.3847/PSJ/ae18da}

\bibitem[{{Proudfoot} {et~al.}(2025{\natexlab{b}}){Proudfoot}, {Holler},
  {Arimatsu}, {Rommel}, {Collyer}, \&
  {Fern{\'a}ndez-Valenzuela}}]{2025PSJ.....6..146P}
{Proudfoot}, B., {Holler}, B.~J., {Arimatsu}, K., {et~al.} 2025{\natexlab{b}},
  \psj, 6, 146, \dodoi{10.3847/PSJ/addd02}

\bibitem[{{Proudfoot} {et~al.}(2025{\natexlab{c}}){Proudfoot}, {Nolthenius},
  {Holler}, {de Souza-Feliciano}, {Rommel}, {Collyer}, {Grundy}, \&
  {Fern{\'a}ndez-Valenzuela}}]{2025ApJ...993L..38P}
{Proudfoot}, B., {Nolthenius}, R., {Holler}, B.~J., {et~al.}
  2025{\natexlab{c}}, \apjl, 993, L38, \dodoi{10.3847/2041-8213/ae1585}

\bibitem[{{Rambaux} \& {Bois}(2004)}]{2004AA...413..381R}
{Rambaux}, N., \& {Bois}, E. 2004, \aap, 413, 381,
  \dodoi{10.1051/0004-6361:20031446}

\bibitem[{{Ribeiro} {et~al.}(2023){Ribeiro}, {Winter}, {Madeira}, \& {Giuliatti
  Winter}}]{2023MNRAS.525...44R}
{Ribeiro}, T., {Winter}, O.~C., {Madeira}, G., \& {Giuliatti Winter}, S.~M.
  2023, \mnras, 525, 44, \dodoi{10.1093/mnras/stad2362}

\bibitem[{{Rodr{\'\i}guez} {et~al.}(2023){Rodr{\'\i}guez}, {Morgado}, \&
  {Callegari}}]{2023MNRAS.525.3376R}
{Rodr{\'\i}guez}, A., {Morgado}, B.~E., \& {Callegari}, Jr., N. 2023, \mnras,
  525, 3376, \dodoi{10.1093/mnras/stad2413}

\bibitem[{{Sicardy} {et~al.}(2025){Sicardy}, {Salo}, {El Moutamid}, {Renner},
  \& {Souami}}]{2025AA...704A..23S}
{Sicardy}, B., {Salo}, H., {El Moutamid}, M., {Renner}, S., \& {Souami}, D.
  2025, \aap, 704, A23, \dodoi{10.1051/0004-6361/202556950}

\end{thebibliography}

\end{document}